\documentclass[journal]{IEEEtran}

\usepackage{amsmath}
\usepackage{graphicx}
\usepackage{threeparttable}
\usepackage{amssymb}
\usepackage{amsthm}
\usepackage{makecell}
\usepackage{subfigure}
\usepackage{cite}
\usepackage{multirow}

\newtheorem{proposition}{Proposition}

\newtheorem{remark}{Remark}
\ifCLASSINFOpdf

\else

\fi

\begin{document}
	
	\title{Learning Precoding in Multi-user Multi-antenna Systems: Transformer or Graph Transformer?}

	\author{Yuxuan~Duan,
		Jia~Guo,
		and~Chenyang~Yang \vspace{-6mm}% <-this % stops a space
		
		\thanks{The authors are with the School of Electronics and Information Engineering, Beihang University, Beijing 100191, China (e-mail: $\{$yuxuanduan, guojia, cyyang$\}$ @buaa.edu.cn).}}
	
	%\markboth{Journal,~Vol.~14, No.~8, July~2024}%
	%{Shell \MakeLowercase{\textit{et al.}}: Bare Demo of IEEEtran.cls for IEEE Journals}

	\maketitle
	\begin{abstract}
		Transformers have been designed for channel acquisition tasks such as channel prediction and other tasks such as precoding, while graph neural networks (GNNs) have been demonstrated to be efficient for learning a multitude of communication tasks. Nonetheless, whether or not Transformers are efficient for the tasks other than channel acquisition and how to reap the benefits of both architectures are less understood. In this paper, we take learning precoding policies in multi-user multi-antenna systems as an example to answer the questions. We notice that a Transformer tailored for precoding can reflect multi-user interference, which is essential for its generalizability to the number of users. Yet the tailored Transformer can only leverage partial permutation property of precoding policies and hence is not generalizable to the number of antennas, same as a GNN learning over a homogeneous graph. To provide useful insight, we establish the relation between Transformers and the GNNs that learn over heterogeneous graphs. Based on the relation, we propose Graph Transformers, namely 2D- and 3D-Gformers, for exploiting the permutation properties of baseband precoding and hybrid precoding policies. The learning performance, inference and training complexity, and size-generalizability of the Gformers are evaluated and compared with Transformers and GNNs via simulations. 	
		
	\end{abstract}
	\begin{IEEEkeywords}
		Transformer, graph neural network, Graph Transformer, precoding, size-generalizability
	\end{IEEEkeywords}

	\IEEEpeerreviewmaketitle
	
	\section{Introduction}\label{Intro}
	\IEEEPARstart{T}{ransformer} has been widely used in the fields of natural language processing (NLP), computer vision, and large language models \cite{achiam2023gpt}.
	Encouraged by its remarkable success, Transformer has been introduced to wireless communications for the tasks such as channel prediction/estimation \cite{CP1,CPTWC,trans_CE_payless_TWC} and channel compression \cite{transCSIfeed3,transCSIfeed4}. It has been proved that the Transformer with an encoder-decoder architecture is equivalent to a recurrent neural network (RNN) \cite{transformer-RNN}, making it a natural fit for channel prediction. Since the attention mechanism can model the correlation between channels in spatial and frequency domains, the Transformer is also powerful for channel estimation and compression.
	
	Recent works have tailored the Transformer architectures for precoding \cite{li2024hpe}.
	%or used graph attention networks (GATs) for learning precoding \cite{GAT_MISO_SE_TWC24}, which can be seen as a variant of Transformer when learning over a homogeneous complete graph.
	Although satisfactory performance can be achieved by the designed architectures, whether or not Transformers are efficient for the wireless tasks other than channel acquisition remains unclear.

	Inspired by the superior generalizability to unseen graph sizes and scalability to large-scale systems, another class of deep neural network (DNN) architectures, graph neural networks (GNNs), has been investigated for a multitude of wireless tasks including power allocation \cite{GNNpc, Letaief, GNNpc5}, link scheduling \cite{linksheduling2}, and precoding \cite{Bipartite, zhao2022understanding, guo2023model, guo2024recursive,liu2023multidimensional,GAT_MISO_SE_TWC24,GNNpc2}.
	It has been realized that these benefits of GNNs stem from harnessing the permutation properties  (say one-dimensional (1D)- or two-dimensional (2D)-permutation equivariance (PE) property) that broadly exist in wireless policies, i.e., the mappings from known parameters (say channel matrix) to decisions (say precoding matrix)  \cite{GNNpc, Letaief, GNNpc5, liu2023multidimensional}.
	
	In order to take advantage of both Transformers and GNNs, Graph Transformers have been proposed recently \cite{GFORMER2024}. However, the rationale for adopting such architectures and how to design them for wireless tasks are still open.
	
	%Recently, another class of deep neural network (DNN) architecture, graph neural networks (GNNs), have also been widely employed in wireless tasks including power allocation \cite{GNNpc, Letaief, GNNpc5}, link scheduling \cite{linksheduling2}, and precoding \cite{zhao2022understanding, guo2023model, guo2024recursive,liu2023multidimensional,GAT_MISO_SE_TWC24,Bipartite, GNNpc2}, . Compared to fully-connected neural networks (FNNs) and convolutional neural networks (CNNs), GNNs exhibit superior generalizability to unseen graph sizes \cite{GAT_MISO_SE_TWC24, Letaief, GNNpc}, scalability to large-scale systems \cite{linksheduling2，GNN-PC-CellFree-TWC2024} and lower training complexity \cite{liu2023multidimensional}.
	%It has been noticed that the advantages of GNNs stem from harnessing permutation properties that widely exist in wireless policies \cite{GNNpc5, liu2023multidimensional}, i.e., the mappings from known parameters (say channel matrix) to decisions (say precoding matrix).
	
	\vspace{-2mm}
	\subsection{Related Works}\vspace{-1mm}
	\subsubsection{Learning Wireless Policies with GNNs}
	GNNs have been extensively investigated for wireless communications, which learn over graphs. In each layer of a GNN, the representation of every vertex or edge in a graph is updated by first extracting information from neighboring vertices or edges with a \emph{processor}, then aggregating the information with a \emph{pooling function}, and finally combining the information with the representation of the vertex or edge by a \emph{combiner}.
	
	The design of a GNN for wireless tasks includes modeling a graph (i.e., defining vertices, edges, and their features) and determining its core components (i.e., combiners, processors, pooling functions, and read-out functions at the output layer).
	
	The graph modeling affects the permutation property satisfied by a GNN  \cite{liu2023multidimensional}.
	The majority of existing works of designing GNNs for learning wireless policies constructed graphs heuristically, say defining a user as a vertex. In this way, the permutability of some dimensions of a policy is often neglected, i.e., the permutation property satisfied by the GNN learning over the graph is not matched with the property of the policy.
	%It has been shown that if the graph is not appropriately modeled, the permutation property satisfied by the GNN is not matched with the property of the wireless policy, leading to the degradation of learning performance and efficiency \cite{GNNpc5}.
	For instance, by modeling users as vertices, the graph attention network (GAT) designed in \cite{GAT_MISO_SE_TWC24} for learning the precoding policy in a multi-user multi-input-single-output (MU-MISO) system (which has user dimension and antenna dimension) is only equivariant to the permutations of users, while the equivariance to the permutations of antennas is overlooked. Since satisfying the permutation property of a dimension is a necessary condition for a DNN to be size-generalizable to the dimension \cite{GNNpc,guo2024recursive}, the GAT is not generalizable to the number of antennas. It is also with high training complexity due to not fully exploiting the PE property of the precoding policy.
	% As a consequence, the GAT is not generalizable to the number of antennas since satisfying the permutation property of a dimension is a necessary condition for a DNN to be size-generalizable to the dimension \cite{GNNpc,guo2024recursive}, and is with high training complexity.
	
	%To avoid property mismatch, parameter-sharing schemes of GNNs were designed in \cite{liu2023multidimensional} to guarantee that the learned policies are with expected permutation properties.
	
	The design of the components in a GNN not only affects its permutation property, but also determines its generalizability to graph sizes. For example, it has been demonstrated in \cite{WJJ_GNN_Gen} that the pooling and read-out functions of GNNs affect the size-generalizability for learning power and bandwidth allocation policies. It has been noticed in \cite{guo2024recursive} that even for a GNN with matched permutation property to a policy, its processor must be designed to reflect the multi-user interference (MUI) in order to enable the size-generalizability to users when learning precoding. Without a judicious design of the components, the GNNs are not size-generalizable, and their training complexity is high.
	
	\subsubsection{Learning Wireless Policies with Transformers}
	The Transformer proposed in \cite{vaswani2017attention} originally for NLP tasks
	has an encoder-decoder architecture. In the encoder, the representation of every token in a sequence is updated over layers, and the order of tokens is distinguished by a positional encoding. In the
	decoder, another sequence is generated. The core component of Transformers is the attention mechanism, which reflects the dependence among the representations of tokens in the sequence.
	Since the dependence can be ``translated'' into correlation, most existing works in the literature of wireless communications designed Transformers for channel prediction, estimation, and compression \cite{CPTWC, trans_CE_payless_TWC, transCSIfeed_novel}.
	
	The design of Transformers for wireless tasks focuses on determining their architectures and defining tokens.
	
	The encoder-decoder architecture was adopted for channel acquisition. In \cite{CPTWC}, a Transformer was designed for channel prediction by exploiting the temporal correlation of channels, where the input and output representations of tokens are respectively the historical and future channels.
	By regarding the pilots and estimated channels as two sequences of tokens,
	a Transformer was designed in \cite{trans_CE_payless_TWC} for channel estimation, by exploiting channel correlation in time and frequency domain.
	In \cite{transCSIfeed_novel}, a Transformer-based architecture was designed for channel feedback to harness channel correlation in the spatial and frequency domain, where the encoder and decoder are respectively used for compressing and recovering channels.
	
	The encoder-only architecture without positional encoding was adopted for other wireless tasks. In \cite{li2024hpe}, such a Transformer was designed to optimize multi-group multicast precoding, where each user was defined as a token.
	The Transformer was actually used to learn a power allocation policy, and the precoding matrix was recovered from the learned powers with an optimal solution structure.
	In \cite{mehrabian2024joint}, such a Transformer was adopted to optimize the bandwidth of subbands, precoding, and reflection coefficients in a reconfigurable intelligent surface-aided multi-antenna system, again each user was defined as a token. It was proved that the tailored Transformer is equivariant to the permutation of users, which was interpreted as the reason for size-generalizability to users.
	
	Different from the tasks of channel acquisition, the rationale of adopting Transformers for other wireless tasks such as precoding was not well-understood so far. Moreover, the Transformers for these tasks were designed heuristically, lacking a guideline for
	designing the architecture, as well as the tokens and their representations.
	\subsubsection{Relation between Transformers and GNNs}\label{sec:intro-relation}
	The connection between the Transformer in \cite{vaswani2017attention} and GNNs has been analyzed in a couple of works. By defining each token as a vertex, the Transformer can be regarded as a GNN that learns over a homogeneous complete graph (called homo-GNN) with only one type of vertices and edges, and every vertex is connected to all other vertices with edges \cite{gradient, GNNandTrans,transformerisGNN2}.
	
	\vspace{-2mm}\subsection{Motivation and Contributions}\vspace{-0.1mm}
	Precoding plays a critical role in achieving high spectral efficiency (SE) of multi-antenna systems. Learning precoding policies with DNNs is promising in providing high SE with much shorter computing latency than numerical algorithms. However, learning these policies efficiently is challenging \cite{zhao2022understanding}, where the DNNs are often with high training complexity and weak size-generalizability.
	%Previous works have validated that Transformers can be used for learning precoding but did not explain why.
	
	In this paper, we strive to answer the following question: \emph{are Transformers the best choice for learning precoding efficiently?} We find that a Transformer tailored for precoding can reflect the MUI that is essential for its generalizability to the number of users, but it does not exhibit the matched permutation property to the precoding policy. Besides, some components in Transformers incur high computational complexity, which however are useless for precoding. This motivates us to answer another question: \emph{how should Transformers be designed to exploit the permutation property with simplified architectures}? Since precoding policies can be learned by GNNs with matched permutation property over heterogeneous graphs, we build the relation between Transformers and GNNs learning over such graphs to answer the questions. This distinguishes our work from \cite{gradient, GNNandTrans,transformerisGNN2} that only analyzed the relation between Transformer and homo-GNN.
	
	Since the precoding policies in different system setups (e.g., MU-MISO or multi-user multi-input-multi-output system) exhibit different permutation properties \cite{zhao2022understanding,liu2023multidimensional}, the Transformers need to be designed to satisfy different properties. Yet designing Transformers to satisfy all possible permutation properties is not the focus of this paper. For easy exposition, we take baseband precoding in MU-MISO systems as an example, and then discuss the extension to hybrid analog and baseband precoding.
	
	Our major contributions are summarized as follows.
	\begin{itemize}
		\item We demonstrate how Transformers should be designed for learning precoding policies efficiently by determining if the encoder-decoder architecture and the positional encoding are necessary as well as by defining tokens and their representations.
		We establish the relations between the Transformers tailored for precoding and the GNNs that learn over heterogeneous graphs (called heter-GNNs). From the relations, we can find that: 1) unlike the heter-GNNs that can exploit the permutation property of the precoding policy, the Transformers can only leverage partial permutation property, and 2) while the update equations of the Transformers are analogous to the update equations of the GNNs that are size-generalizable to users, their ways of reflecting the MUI differ.
		\item Inspired by the relations, we propose Graph Transformers that can learn over heterogeneous graphs with matched permutation properties to precoding policies, namely 2D-Gformer and 3D-Gformer. We notice that the designed token representations for learning over heterogeneous graphs are quite different from those learning over homogeneous graphs, where a token may even have multiple representations. 
		% The learning performance, inference and training complexity, and size-generalizability of the proposed Gformers are evaluated via simulations and compared with Transformers and GNNs.
		\item We evaluate the training complexity and the size-generalizability of the proposed Gformers via simulations. For learning a SE-maximal baseband policy, it is demonstrated that the 2D-Gformer exhibits lower training complexity and better size-generalizability than the Transformers. For learning a SE-maximal hybrid precoding policy, the 3D-Gformer can be well-generalized to the numbers of users, antennas and RF chains.
	\end{itemize}
	
	The rest of this paper is organized as follows. In section \ref{sec:PE}, we introduce the permutation property of the baseband precoding policy in MU-MISO systems. In section \ref{sec:gnns}, we recap several GNNs proposed for precoding. In section \ref{sec:Transformers}, we show how to tailor the Transformer for learning precoding, and establish the relation between Transformers and GNNs. In section \ref{sec:2D-Gformer}, we propose Graph Transformers for learning baseband precoding and extend to hybrid precoding. In section \ref{sec:simulation}, simulation results are provided. In section \ref{sec:conclusion}, we provide conclusion remarks.

	\textit{Notations}:
	%Vectors and matrices are denoted as lower-case bold letters $\mathbf{x}$ and upper-case bold letters $\mathbf{X}$, respectively.
	$\mathbf{X}=[x_{ij}]$ denotes a matrix with the element in the $i$-th row and $j$-th column being $x_{ij}$, $(\mathbf{X})_{ij}$ denotes the element in the $i$-th row and $j$-th column of $\mathbf{X}$.  $\mathbf{x}_i$ represents the $i$-th column vector of $\mathbf{X}$. $\mathbf{I}$ denotes an identity matrix. $\mathrm{diag}(\cdot)$ denotes a diagonal matrix. $\mathsf{Tr}(\cdot)$ denotes the trace  of a matrix. $(\cdot)^{\mathsf{T}}$ and $(\cdot)^{\mathsf{H}}$ denote the transpose and conjugate transpose of a matrix or vector, respectively. $\mathsf{Re}(\cdot)$ and $\mathsf{Im}(\cdot)$ respectively stand for the real and the imaginary part of a complex value.
	
	%$\Vert \cdot \Vert_F$ denotes the Frobenius norm
	
	\section{Baseband Precoding and Permutation Property}\label{sec:PE}
	Consider a downlink MU-MISO system, where a base station (BS) equipped with $N$ antennas serves $K$ single-antenna users.
	
	The precoding at the BS can be designed from an optimization problem, say SE-maximization problem under power constraint or energy-efficient maximization problem under both transmit power and quality of service constraints.
	%say the SE-maximization problem as follows,
	%\begin{subequations}\label{Sumrate Prob}
	%	\begin{align}
		%		\mathop{\max}_{\mathbf{V}}\quad\quad &\sum_{k=1}^K \mathrm{log_2} \left(1 + \frac{|\mathbf{h}_k^\mathsf{H} \mathbf{v}_k|^2}{\sum_{j=1,j \neq k}^{K} |\mathbf{h}_k^{\mathsf{H}} \mathbf{v}_j|^2 + \sigma_0^2} \right), \\
		%		\mathrm{s.t.} \quad\quad &\mathsf{Tr}\left(\mathbf{V}^\mathsf{H} \mathbf{V} \right) \leq P_t, \label{Sumrate Constraint}
		%	\end{align}
	%\end{subequations}
	%where $\mathbf{h}_k \triangleq [h_{k1},...,h_{kN}] \in \mathbb{C}^{N}$ and $\mathbf{v}_k \triangleq [v_{k1},...,v_{kN}] \in \mathbb{C}^{N}$ are respectively the channel and precoding vectors of the $k$-th user, $\mathbf{H} \triangleq \begin{bmatrix}
		%	\mathbf{h}_1,...,\mathbf{h}_K
		%\end{bmatrix} \in \mathbb{C}^{N \times K}$ and $\mathbf{V} \triangleq \begin{bmatrix}
		%	\mathbf{v}_1, ...,\mathbf{v}_K
		%\end{bmatrix} \in \mathbb{C}^{N \times K}$ are respectively the channel and precoding matrices, $P_t$ and $\sigma_0^{2}$  are respectively the maximal transmit power at the BS and the noise power.
		
		The precoding policy is a mapping from the channel matrix $\mathbf{H}$ to the optimal precoding matrix $\mathbf{V}^*$, which is denoted as
		\begin{equation}\label{precoding policy}
			\mathbf{V}^*=g_v(\mathbf{H}),
		\end{equation}
		where $\mathbf{H} = [h_{nk}] \in \mathbb{C}^{N \times K}$, $\mathbf{V}^* =[v_{nk}^*] \in \mathbb{C}^{N \times K}$, and $h_{nk}$ is the channel coefficient from the $n$-th antenna to the $k$-th user.
		
		It has been proved in \cite{zhao2022understanding} that the precoding policy satisfies the following 2D-PE property,
		\begin{equation}
			\label{2d-pe}
			\mathbf{\Pi}_{\mathsf{AN}}\mathbf{V}^* \mathbf{\Pi}_{\mathsf{UE}} = g_v(\mathbf{\Pi}_{\mathsf{AN}}\mathbf{H} \mathbf{\Pi}_{\mathsf{UE}}),
		\end{equation}
		where $\mathbf{\Pi}_{\mathsf{AN}}$ and $\mathbf{\Pi}_{\mathsf{UE}}$
		are permutation matrices that change the orders of antennas and users, respectively.
		
		When $\mathbf{\Pi}_{\mathsf{AN}} = \mathbf{I}$ or $\mathbf{\Pi}_{\mathsf{UE}} = \mathbf{I}$, the 2D-PE property is degenerated into the following 1D-PE properties,
		\begin{equation}
			\label{1d-pe-UE}
			\mathbf{V}^*\mathbf{\Pi}_{\mathsf{UE}}  = g_v(\mathbf{H}\mathbf{\Pi}_{\mathsf{UE}}),
		\end{equation}
		\begin{equation}
			\label{1d-pe-AN}
			\mathbf{\Pi}_{\mathsf{AN}}\mathbf{V}^*  = g_v(\mathbf{\Pi}_{\mathsf{AN}}\mathbf{H}).
		\end{equation}
		If  \eqref{2d-pe} holds, then \eqref{1d-pe-UE} or \eqref{1d-pe-AN} must hold. If both \eqref{1d-pe-UE} and \eqref{1d-pe-AN} hold, then \eqref{2d-pe} is true.
		
		For the SE-maximization problem subject to the constraint of total power at the BS $P_\mathrm{t}$, the optimal baseband precoding matrix has the following structure \cite{Björnson},
		\begin{IEEEeqnarray}{rcl}
			\mathbf{V}^*=\left(\mathbf{I}+\frac{1}{\sigma^2}\mathbf{H}\boldsymbol{\Lambda}\mathbf{H}^\mathsf{H}\right)^{-1}\mathbf{H}\mathbf{P}^\frac{1}{2},
			\label{eq_duality}
		\end{IEEEeqnarray}
		\noindent where $\mathbf{P}=\mathrm{diag}(p_1/||(\mathbf{I}+\frac{1}{\sigma^2}\mathbf{H}\boldsymbol{\Lambda}\mathbf{H}^\mathsf{H})^{-1}\mathbf{h}_1||^2,\cdots,$ $p_K/||(\mathbf{I}+\frac{1}{\sigma^2}\mathbf{H}\boldsymbol{\Lambda}\mathbf{H}^\mathsf{H})^{-1}\mathbf{h}_K||^2)$, $\boldsymbol{\Lambda}=\mathrm{diag}(\lambda_1,\cdots,\lambda_K)$, $p_1,\cdots,p_K$ are the allocated powers, $\lambda_k>0, k=1,\cdots,K$, $\sum_{k=1}^K\lambda_k=\sum_{k=1}^K{p_k}=P_\mathrm{t}$, and $\sigma^2$ is the noise power.
		
		With the structure, the precoding matrix can also be obtained indirectly by first learning a power allocation policy (denoted as $(\mathbf{p}^*,\boldsymbol{\lambda}^*)=g_p(\mathbf{H})$) and then recovering from \eqref{eq_duality}, instead of learning the precoding policy directly. This is a popular model-driven deep learning method for SE-maximal baseband precoding \cite{xia2020deep,Bipartite,ZBCGC2023}, because the power allocation policy $g_p(\cdot)$ is much easier to learn than the precoding policy $g_v(\cdot)$ \cite{zhao2022understanding}.
		
		% , which satisfies 1D-PE property, i.e., $(\boldsymbol{\Pi}_{\mathsf{UE}}^\mathsf{T}\mathbf{p}^*,\boldsymbol{\Pi}_{\mathsf{UE}}^\mathsf{T}\boldsymbol{\lambda}^*)=g_p(\boldsymbol{\Pi}_{\mathsf{UE}}^\mathsf{T}\mathbf{H})$\cite{ZBCGC2023}
		
		\section{GNNs for the Precoding Policy}\label{sec:gnns}
		In this section, we briefly review several GNNs designed for learning the baseband precoding policy, which are with matched PE property to the precoding policy.
		
\begin{figure}
			\centering
			\includegraphics[width=0.7\linewidth]{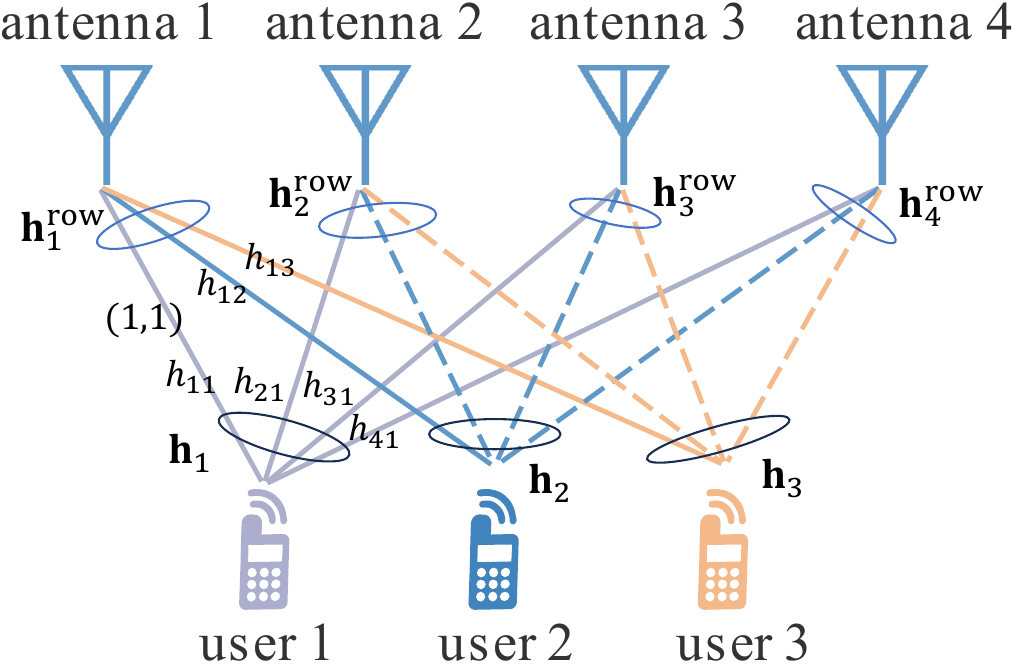}\vspace{-2mm}
			\caption{Illustration of the modeled heterogeneous graph for precoding, $N=4$, $K=3$. The neighboring edges of $(1,1)$ are drawn with solid lines.}
			\label{Graph}
		\end{figure}

		To learn the precoding policy by GNNs for satisfying the 2D-PE property, a heterogeneous graph was constructed in \cite{zhao2022understanding}. As illustrated in Fig. \ref{Graph}, the graph has two types of vertices: antenna vertices and user vertices.
		Each user vertex is connected with each antenna vertex by an edge.
		There are no features or actions on vertices. The feature and action of edge $(n,k)$, i.e., the edge connecting the $n$-th antenna vertex and the $k$-th user vertex, are respectively $h_{nk}$ and $v_{nk}$. The edges are neighbored if they are connected to the same vertex. For example, edge $(i,k), i\neq n$ and edge $(n,j), j\neq k$ are the neighboring edges of edge $(n,k)$.
		
		Since all the features and actions are defined on edges, the precoding policy can be learned with Edge-GNNs,  where the representation of every edge is updated.  The representation of edge $(n,k)$ in the $\ell$-th layer is denoted as $\mathbf{d}_{\mathsf{E},nk}^{(\ell)} \in \mathbb{R}^{J^{(\ell)}}$, where $J^{(\ell)}$ is the dimension of the representation.
		To update $\mathbf{d}_{\mathsf{E},nk}^{(\ell)}$, the information from the neighboring edges of edge $(n,k)$ is first extracted with a \emph{processor} and aggregated with a \emph{pooling function}, and then combined with $\mathbf{d}_{\mathsf{E},nk}^{(\ell-1)}$ by a \emph{combiner}.\footnote{The precoding policy can also be learned with a Vertex-GNN, where the representation of every vertex is updated, and the fully-connected neural networks (FNNs) are adopted as processors \cite{WeiYu,vertexGNNorEdgeGNN}. If the GNN learns over the graph in Fig. \ref{Graph} for satisfying the 2D-PE property, an extra read-out layer is required to convert the vertex representations into the actions on edges.  Many existing works use Vertex-GNNs to optimize precoding in various settings directly or indirectly over a homogeneous graph only with user vertices \cite{GAT_MISO_SE_TWC24,WeiYu,GAT_MISO_EE,GAT_Hybrid}, which only exhibit the 1D-PE property in \eqref{1d-pe-UE}.}
		The representations of edge $(n,k)$ in the first layer and last layer are $\mathbf{d}_{\mathsf{E},nk}^{(0)} \triangleq [\mathsf{Re}(h_{nk}),\mathsf{Im}(h_{nk})]^\mathsf{T}$ and $\mathbf{d}_{\mathsf{E},nk}^{(L)} \triangleq [\mathsf{Re}(v_{nk}),\mathsf{Im}(v_{nk})]^\mathsf{T}$ (i.e., $J^{(0)}=J^{(\ell)}=2$), respectively, where $L$ is the number of layers.
		
		The architecture of a GNN can be expressed by its update equation, and the difference among the Edge-GNNs lies in the processors, pooling functions, and combiners.
		In what follows, we recap several Edge-GNNs proposed for precoding in the literature. To facilitate the analyses of their relation with the Transformer later, their update equations are expressed in the form of updating representations of all the edges connected to each user vertex (say for the $k$-th user vertex in the $\ell$-th layer, denoted as $\mathbf{d}_{\mathsf{E},k}^{(\ell)}\triangleq[\mathbf{d}_{\mathsf{E},1k}^{(\ell)\mathsf{T}},\cdots,\mathbf{d}_{\mathsf{E},Nk}^{(\ell)\mathsf{T}}]^{\mathsf{T}} \in \mathbb{R}^{NJ^{(\ell)}}$), instead of updating the representation of each edge.
		
		\vspace{-3mm}
		\subsection{Edge-GCN}\label{sec:vgnn} The graph convolutional network (GCN) uses a linear processor and sum pooling function, whose combiner is a linear function cascaded by an activation function $\sigma(\cdot)$. The update equation in the $\ell$-th layer of the Edge-GCN designed in \cite{zhao2022understanding} can be expressed as,
		\begin{equation}
			\label{vanilla update Eq}
			\begin{aligned}
				\mathbf{d}_{\mathsf{E},k}^{(\ell)}&=\sigma\Big(\mathbf{W}\mathbf{d}_{\mathsf{E},k}^{(\ell-1)}+\sum_{i=1}^K\mathbf{U}\mathbf{d}_{\mathsf{E},i}^{(\ell-1)}\Big),
				\\&\triangleq f_{\mathsf{C}}\left(\mathbf{d}_{\mathsf{E},k}^{(\ell-1)},\sum_{i=1}^{K}q_{\mathsf{C}}\left(\mathbf{d}_{\mathsf{E},i}^{(\ell-1)}\right)\right),
			\end{aligned}
		\end{equation}
		where $\mathbf{W}$ and $\mathbf{U}$ are trainable weight matrices, and $f_{\mathsf{C}}(\cdot)$ and $q_{\mathsf{C}}(\cdot)$ are respectively the parameterized combiner and linear processor. While $\mathbf{W}$ and $\mathbf{U}$ differ among layers, the superscript $(\ell)$ in the two matrices is omitted for notational simplicity.
		
		When the Edge-GCN learns over the graph illustrated in Fig. \ref{Graph}, $\mathbf{W} \in \mathbb{R}^{NJ^{(\ell)} \times NJ^{(\ell-1)}}$ and $\mathbf{U}\in \mathbb{R}^{NJ^{(\ell)} \times NJ^{(\ell-1)}} $, each is with  $N \times N$ blocks where the diagonal blocks and off-diagonal blocks are respectively identical. Specifically, the two matrices are with the following structure of parameter-sharing,
		\begin{equation}\label{eq:matrix structure}
			\mathbf{W} \!=\! \begin{bmatrix}
				\mathbf{W}_1 &\mathbf{W}_2  &\cdots &\mathbf{W}_2\\
				\mathbf{W}_2  &\mathbf{W}_1  &\cdots & \mathbf{W}_2\\
				\vdots & \vdots & \ddots  &\vdots\\
				\mathbf{W}_2  &\mathbf{W}_2  &\cdots  &\mathbf{W}_1
			\end{bmatrix},
			\mathbf{U} \!=\!  \begin{bmatrix}
				\mathbf{U}_1 &\mathbf{U}_2  &\cdots &\mathbf{U}_2\\
				\mathbf{U}_2  &\mathbf{U}_1  &\cdots & \mathbf{U}_2\\
				\vdots & \vdots & \ddots  &\vdots\\
				\mathbf{U}_2  &\mathbf{U}_2  &\cdots  &\mathbf{U}_1
			\end{bmatrix}.
		\end{equation}

		\vspace{-3mm}
		\subsection{RGNN} The recursive GNN (RGNN) proposed in  \cite{guo2024recursive} satisfies the high dimensional PE property by satisfying the PE property of each dimension in a recursive manner. Its update equation in the $\ell$-th layer can be expressed as,
		\begin{equation}
			\label{RGNN update eq}
			\mathbf{d}_{\mathsf{E},k}^{(\ell)}
			=f_{\mathsf{R}}\left(\mathbf{d}_{\mathsf{E},k}^{(\ell-1)},\sum_{i=1}^{K} q_{\mathsf{R}}\left(\mathbf{d}_{\mathsf{E},k}^{(\ell-1)},\mathbf{d}_{\mathsf{E},i}^{(\ell-1)}\right)\right),
		\end{equation}
		which satisfies the PE property of user dimension (i.e., the 1D-PE property in \eqref{1d-pe-UE}), where both the combiner $f_\mathsf{R}(\cdot)$ and the processor $q_\mathsf{R}(\cdot)$ are parameterized non-linear functions.
		
		When the RGNN learns over the graph in Fig. \ref{Graph}, $f_\mathsf{R}(\cdot)$ or  $q_\mathsf{R}(\cdot)$ is the mapping from $\mathbf{x}$ to $\mathbf{y}$  that satisfies the 1D-PE property in \eqref{1d-pe-AN}. Each of them is with the form of $y_n=\phi(x_n,\sum_{i=1}^N \psi(x_n, x_i))$, where $\phi(\cdot)$ and $\psi(\cdot)$ are FNNs, $x_n$ and $y_n$ are the $n$-th element in $\mathbf{x}$ and $\mathbf{y}$, respectively.
		
		Different from the Edge-GCN, the processor of the RGNN $q_{\mathsf{R}}(\cdot)$ enables to model the relation between $\mathbf{d}_{\mathsf{E},k}^{(\ell-1)}$ and $\mathbf{d}_{\mathsf{E},i}^{(\ell-1)}$ with training. Hence, the RGNN is with attention mechanism. Since $\mathbf{d}_{\mathsf{E},k}^{(\ell-1)}$ and $\mathbf{d}_{\mathsf{E},i}^{(\ell-1)}$ are the representations of the edges respectively connected to the $k$-th user vertex and the $i$-th user vertex, the processor can model the correlation between the channel vectors of the two users and thereby implicitly \emph{reflect the MUI} between them.
		
		% $\mathbf{d}_{\mathsf{E},k}^{(0)}=[\mathsf{Re}(\mathbf{h}_k^{\mathsf{H}}),\mathsf{Im}(\mathbf{h}_k^{\mathsf{H}})]^{\mathsf{H}}$ and $\mathbf{d}_{\mathsf{E},i}^{(0)}=[\mathsf{Re}(\mathbf{h}_i^{\mathsf{H}}),\mathsf{Im}(\mathbf{h}_i^{\mathsf{H}})]^{\mathsf{H}}$
		
		\vspace{-3mm}\subsection{Model-GNN}\label{sec: Model-GNN} In contrast to the Edge-GCN and the RGNN that are purely data-driven, a mathematical model of Taylor's expansion of matrix pseudo-inverse is incorporated into the Model-GNN proposed in \cite{guo2023model} for learning the precoding policy.
		
		In \cite{guo2023model}, each layer of the  Model-GNN consists of a TGNN sub-layer that comes from the Taylor's expansion, followed by an Edge-GCN sub-layer with the update equation as in \eqref{vanilla update Eq}. For better comparison, we omit the information aggregation step of the Edge-GCN sub-layer, which does not affect the learning performance according to our evaluation. Then, the update equation in the $\ell$-th layer of the simplified Model-GNN can be expressed as,
		\begin{equation}
			\begin{aligned}
				\mathbf{d}_{\mathsf{E},k}^{(\ell)} &= \sigma\Big(\mathbf{W} \mathbf{d}_{\mathsf{E},k}^{(\ell-1)} + \mathbf{U} \sum_{i=1}^{K}
				\underbrace{\left(\mathbf{d}_{\mathsf{E},k}^{(\ell-1){\mathsf{H}}} \mathbf{h}_{i}\right)}_{\alpha_{ki}^{\mathsf{M}}}\mathbf{d}_{\mathsf{E},i}^{(\ell-1)}\Big) \\& \triangleq f_{\mathsf{M}}\left(\mathbf{d}_{\mathsf{E},k}^{(\ell-1)},\sum_{i=1}^Kq_\mathsf{M}\left(\mathbf{d}_{\mathsf{E},k}^{(\ell-1)},\mathbf{h}_{i},\mathbf{d}_{\mathsf{E},i}^{(\ell-1)}\right)\right),\label{Model-GNN update eq}
			\end{aligned}
		\end{equation}
		where $\alpha_{ki}^{\mathsf{M}} \triangleq \mathbf{d}_{\mathsf{E},k}^{(\ell-1){\mathsf{H}}} \mathbf{h}_{i}$ is the attention score that can reflect the MUI between the $k$-th and $i$-th users, and the term $\sum_{i=1}^{K}\alpha_{ki}^{\mathsf{M}}\mathbf{d}_{\mathsf{E},i}^{(\ell-1)}$ comes from the Taylor's expansion. The combiner $f_\mathsf{M}(\cdot)$ is a parameterized non-linear function, while the processor $q_\mathsf{M}(\cdot)$ is a non-linear function of $\mathbf{d}_{\mathsf{E},k}^{(\ell-1)}$ that computes the attention score between $\mathbf{d}_{\mathsf{E},k}^{(\ell-1)}$ and $\mathbf{h}_i$ without trainable parameters.
		% $\mathbf{d}_{\mathsf{E},k}^{(\ell-1)}$ and $\sum_{i=1}^{K}\alpha_{ki}\mathbf{d}_{\mathsf{E},i}^{(\ell-1)}$ are weighted with two trainable weight matrices $\mathbf{U}$ and $\mathbf{W}$ in the combiner, such that the direction and power of $\mathbf{d}_{\mathsf{E},k}^{(\ell)}$ can be adjusted to find the optimal precoding vector \cite{guo2023model}.
		The matrices $\mathbf{W}$ and $\mathbf{U}$ are also with the structure in \eqref{eq:matrix structure}.
		
		\vspace{-2mm}
		\begin{figure}[!htbp]
			\centering
			% \subfigure[Overall architecture of the GNNs for learning precoding]{
				% 	\includegraphics[width=0.19\linewidth]{Figures/a1.png}\label{overall}}
			\subfigure[Edge-GCN]{
				\includegraphics[width=0.31\linewidth]{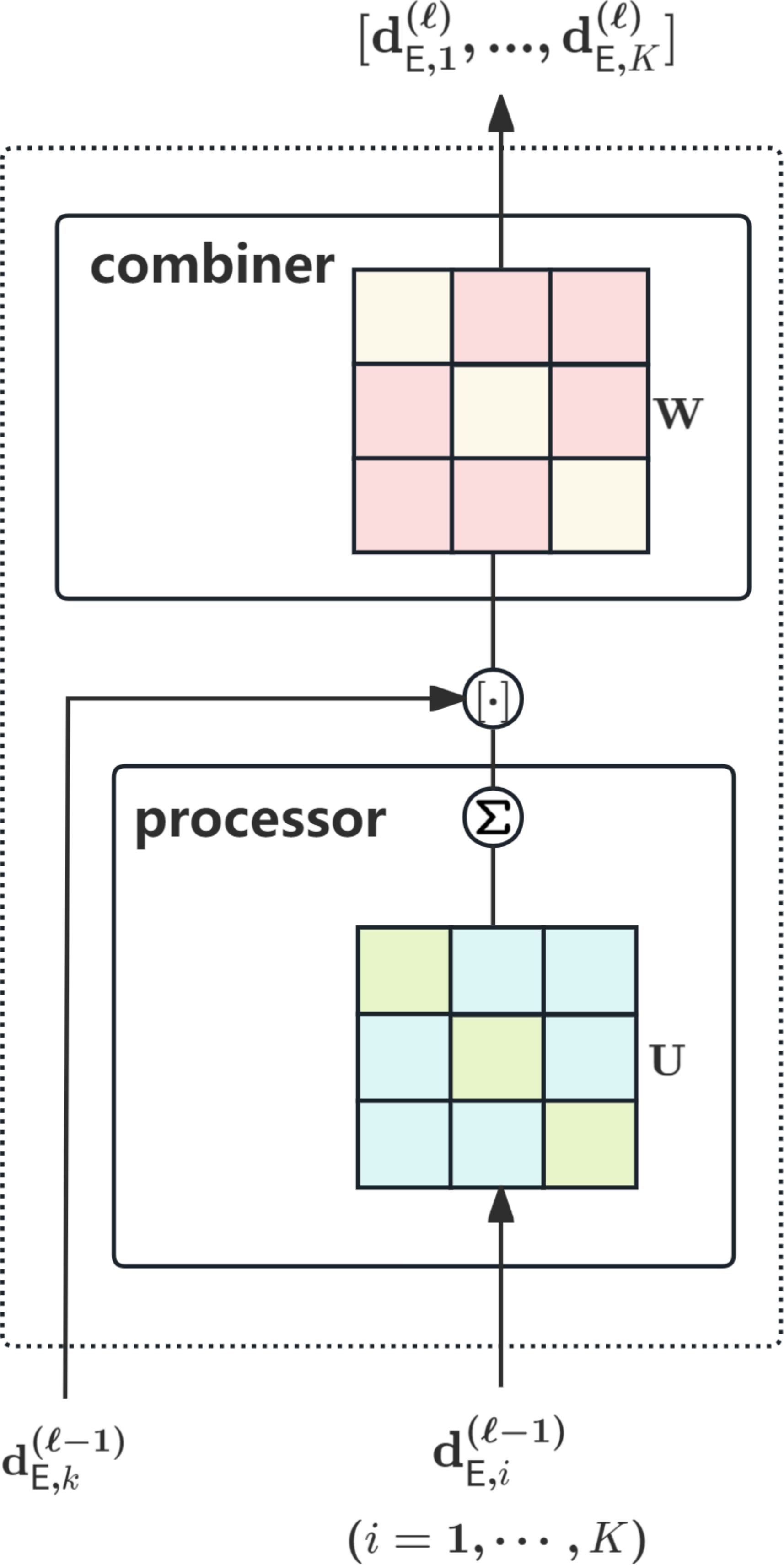}\label{fig: EdgeGCN}}\subfigure[RGNN]{
				\includegraphics[width=0.31\linewidth]{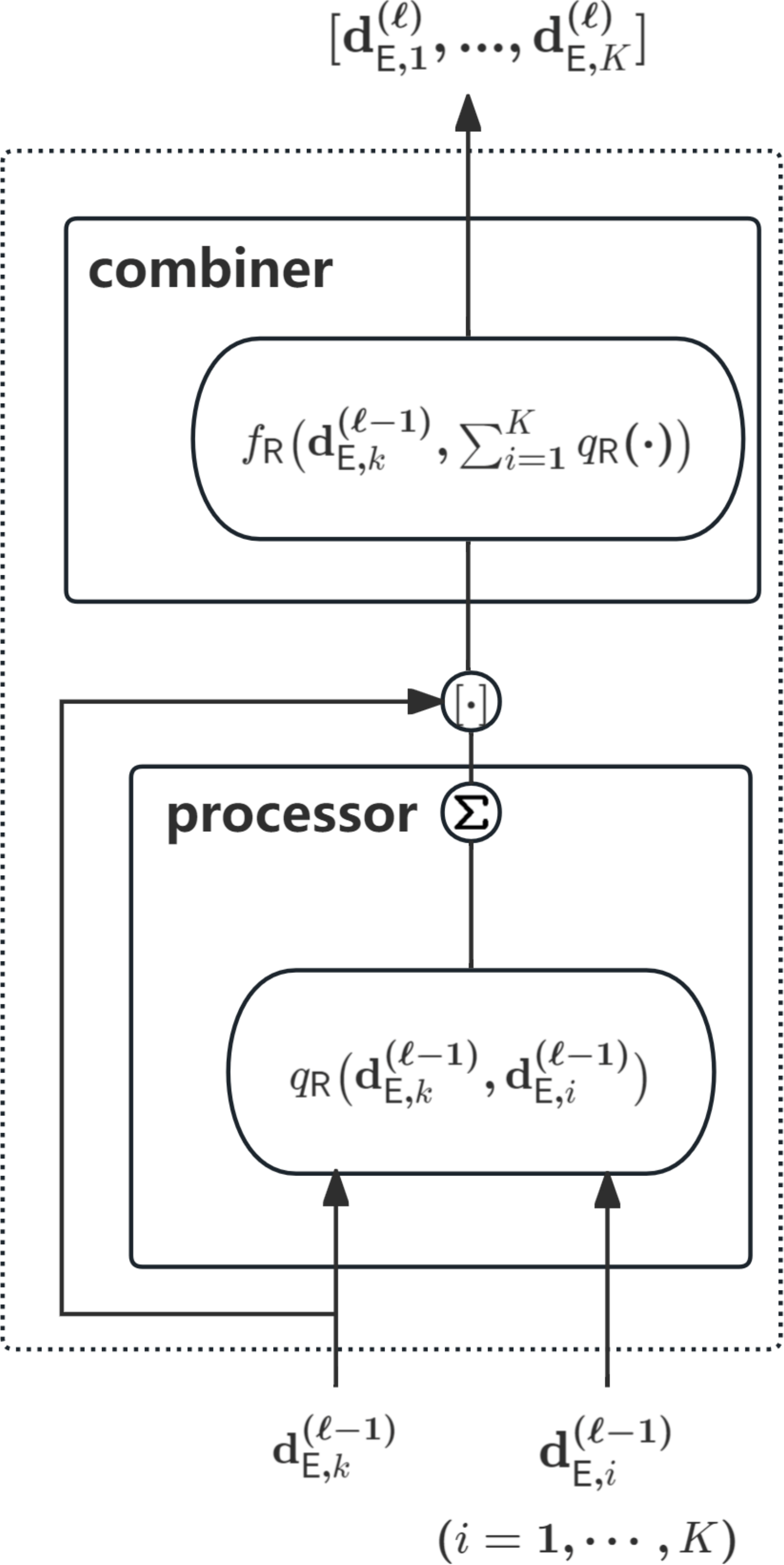}\label{fig: RGNN}}\hfil\subfigure[Model-GNN]{
				\includegraphics[width=0.31\linewidth]{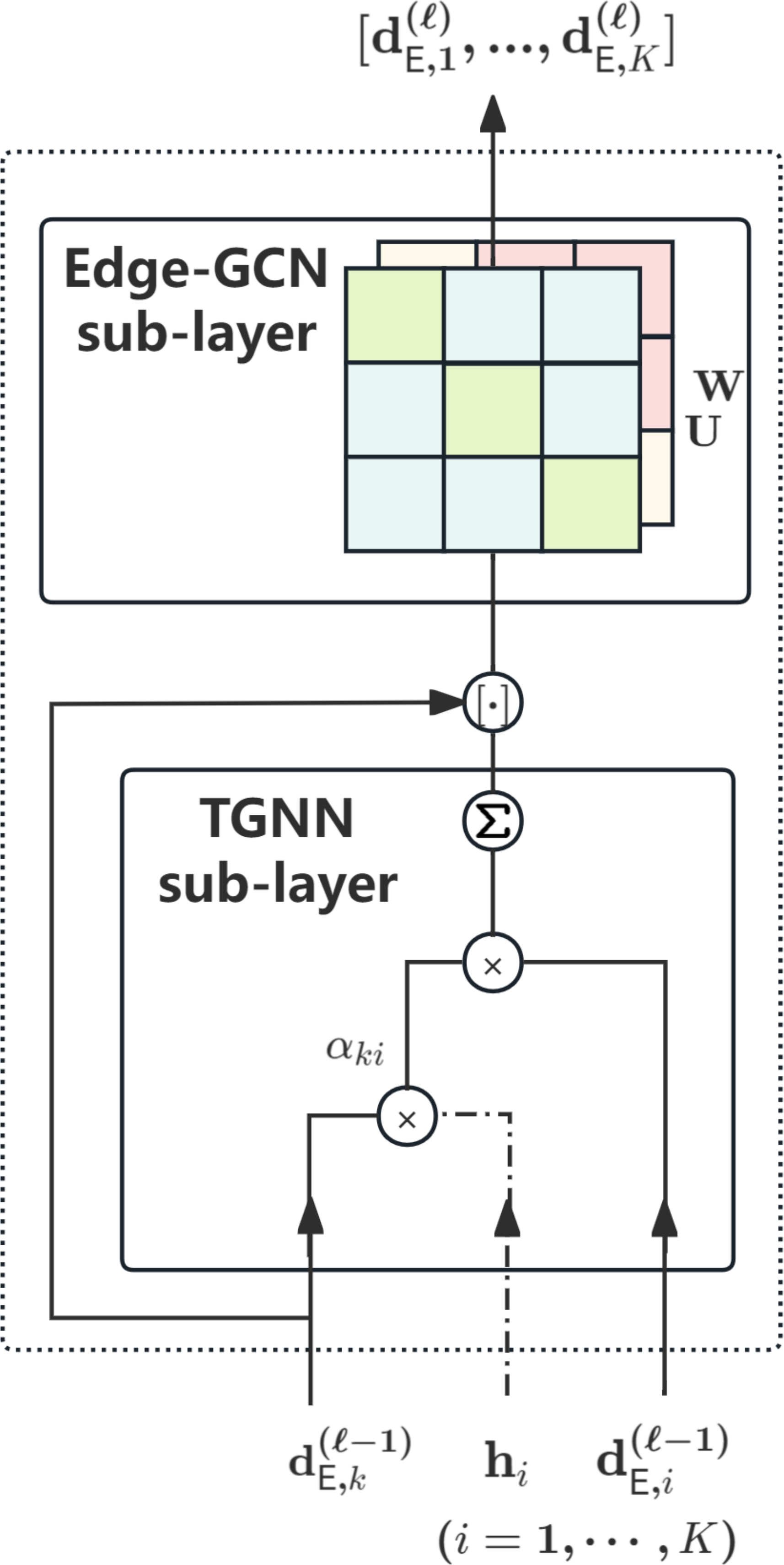}\label{fig: ModelGNN}}
			\vspace{-2mm}
			\caption{Architecture of the $\ell$-th layer in each Edge-GNN. Identical matrix blocks are represented using the same color.}\label{overall}
		\end{figure}\vspace{-2mm}

		The architectures of these Edge-GNNs are shown in Fig. \ref{overall}.
		
		% By comparing \eqref{vanilla update Eq} and \eqref{Model-GNN update eq}, we can see that the processor of the Model-GNN differs from that in the Edge-GCN. Specifically, $\big[\mathbf{d}_{\mathsf{E},i}^{(\ell-1)}\big]$ are weighted equally by $\mathbf{U}$ with the Edge-GCN, but are weighted differently by $\alpha_{ki}$ with the Model-GNN (shown in \eqref{model sub1}). Since $q_\mathsf{M}(\cdot)$ is a non-linear function of the representations of edges connected to the $k$-th user and the $i$-th user, the processor of Model-GNN is said to \emph{model the interference between the $k$-th user and the $i$-th user}. Such characteristic is crucial for learning precoding policies with low training complexity and size-generalizability \cite{guo2023model}.
		
		It has been shown in \cite{zhao2022understanding,guo2023model,guo2024recursive} that the Edge-GCN, RGNN, and Model-GNN satisfy the PE property of both user dimension and antenna dimension (i.e., satisfy the 2D-PE property in \eqref{2d-pe}), and hence have the potential of generalizing to both numbers of antennas and users. As evaluated therein, these Edge-GNNs can be well-generalized to the number of antennas, whereas only the RGNN and Model-GNN are generalizable to the number of users.
		This is because the processor in GNNs for precoding should reflect the MUI for enabling the size-generalizability to users \cite{guo2024recursive}.
		
		\section{Transformers for the Precoding Policy}\label{sec:Transformers}
		In this section, we first recap the vanilla Transformer proposed in \cite{vaswani2017attention}.
		Then, we explain how to tailor the architecture of the Transformer, and determine the tokens and their representations for learning the precoding policy. By expressing its architecture with update equation, we can analyze the relation between the tailored Transformer and the Edge-GNNs, which facilitates the design of Graph Transformers.

		\vspace{-2mm}\subsection{Vanilla Transformer}
		The vanilla Transformer proposed in \cite{vaswani2017attention} for NLP tasks is with an \emph{encoder-decoder} architecture. The encoder is used for representing a sequence of words in a sentence, and the decoder is used for generating another sequence of words.
		
		When the Transformer is used for machine translation, a sentence of words is first mapped into a sequence of token representations by word embedding. To distinguish the order of the words, a \emph{positional encoding} is added to the token representations, which are then updated through $L$ layers in the encoder. Each layer is composed of an \emph{attention sub-layer} and a \emph{feed-forward network (FFN)} sub-layer. Finally, the output of the encoder is inputted into the decoder to generate the words in the translated sentence.
		
		\vspace{-2mm}\subsection{Transformer for Precoding}\label{1D-Transformers}
		We first present the tailored architecture for precoding, and then introduce two ways of defining the tokens for precoding, where the difference lies in whether the MUI can be reflected.
		
		\emph{\bf Is the encoder-decoder architecture required?}
		A DNN for the baseband precoding strives to learn the mapping from $\mathbf{H}$ to $\mathbf{V}^*$, without the need for sequence generation. Hence, the decoder of the Transformer is useless for learning the policy.
		
		Besides, since $\mathsf{Re}(\mathbf{H})$ and $\mathsf{Im}(\mathbf{H})$ can serve as the input of the DNN, the word embedding is unnecessary.
		%For notational simplicity, we assume that $(\mathbf{H})_{nk}, n=1,\cdots, N, k=1,\cdots, K$ are real values, and denote each element in the output vector of the DNN as $(\mathbf{V})_{nk}, n=1,\cdots, N, k=1,\cdots, K$ by ignoring the normalization for satisfying power constraint in the following analyses.
		
		\emph{\bf Is positional encoding required?}
		For NLP tasks such as machine translation, positional encoding is critical since the order of tokens determines the meaning of a sentence. The Transformer with positional encoding is not equivariant to the permutations of tokens \cite{lin2021survey}. Therefore, the positional encoding should be removed when learning a wireless policy such as precoding, since satisfying the PE property of a dimension is necessary for a DNN to be size-generalizable to the dimension.
		
		%Such an encoder-only Transformer without the positional encoding can be regarded as a GNN that learns over a homogeneous complete graph, where each token corresponds to each vertex.
		%This can be seen by expressing the architecture of the this Transformer by its update equation for each token in each layer.
		The architecture of such an encoder-only Transformer without the positional encoding can be expressed by its update equation for each token representation in each layer. For notational simplicity, only single-head attention is considered, but the following analyses can be easily extended to the Transformers with multi-head attention.
		
		Denote the number of tokens as $N_{\mathsf{token}}$.
		
		\subsubsection{Attention sub-layer}
		In the $\ell$-th layer, the input of the attention sub-layer is the representations of $N_{\mathsf{token}}$ tokens. For the representation of the $k$-th token, denoted as $\mathbf{d}_{k}^{(\ell)}$, the output of the attention sub-layer can be expressed as,
		\begin{equation}\label{eq:1d-attention}
			\mathbf{c}_{k}^{(\ell)}
			=\sum_{i=1}^{N_{\mathsf{token}}}  \xi\Big(\big(\underbrace{\mathbf{W}^{\mathsf{Q}} \mathbf{d}_k^{(\ell-1)}}_{\mathrm{Query}}\big)^{\mathsf{T}} \underbrace{\big(\mathbf{W}^{\mathsf{K}} \mathbf{d}_i^{(\ell-1)}\big)}_{\mathrm{Key}}\Big)\underbrace{\big(\mathbf{W}^{\mathsf{V}} \mathbf{d}_i^{(\ell-1)} \big)}_{\mathrm{Value}},
		\end{equation}
		where $\xi(\cdot)$ represents the softmax function (i.e., $y_k=\frac{\exp(x_k)}{\sum_{j=1}^{N_\mathsf{token}}\exp(x_j)} $), and $\mathbf{W}^{\mathsf{Q}}$, $\mathbf{W}^{\mathsf{K}}$, $\mathbf{W}^{\mathsf{V}}$ are trainable matrices for projecting the token representations. Again, while they differ among layers, we omit the superscript $(\ell)$ in these matrices for notational simplicity. The projected representations are referred to as Query, Key, and Value, respectively \cite{vaswani2017attention}.
		
		The dependence between the representations of the $k$-th and $i$-th tokens is reflected by an attention score, which is
		\begin{equation}\label{eq:1d-attention-original}
			\alpha_{ki}\triangleq \xi\Big(\big(\mathbf{W}^{\mathsf{Q}} \mathbf{d}_k^{(\ell-1)}\big)^{\mathsf{T}} \big(\mathbf{W}^{\mathsf{K}} \mathbf{d}_i^{(\ell-1)}\big)\Big).
		\end{equation}

		% \footnote{There may be a little misuse of the notations because $\mathbf{d}_k^{(\ell)}$ denotes  the representation of the $k$-th user vertex in section \ref{sec:gnns}.}
		
		\subsubsection{FFN sub-layer}
		To avoid gradient vanishing during training, the output of the attention sub-layer $\mathbf{c}_k^{(\ell)}$ is added with $\mathbf{d}_k^{(\ell-1)}$ (i.e., using the residual connection as stated in \cite{vaswani2017attention}) before inputting into the FFN  sub-layer.
		In \cite{vaswani2017attention}, an FNN was adopted as the FFN sub-layer. For notational simplicity, we assume that this FNN is with a single layer, then the output of the FFN sub-layer is,
		\begin{equation}\label{eq:ffn}
			\mathbf{d}_{k}^{(\ell)} = \sigma\left(\mathbf{W}^{\mathsf{F}} \left(\mathbf{d}_{k}^{(\ell-1)}+\mathbf{c}_{k}^{(\ell)}\right)\right),
		\end{equation}
		where $\mathbf{W}^{\mathsf{F}}$ is a trainable weight matrix.
		
		By substituting \eqref{eq:1d-attention} into \eqref{eq:ffn}, the \emph{update equation} for the $k$-th token in the $\ell$-th layer of the tailored Transformer is,
		\begin{eqnarray}\label{eq:upd-transformer}
			\mathbf{d}_{k}^{(\ell)} \!\!&\!\!=\!\!&\!\! \sigma\Bigg(\mathbf{W}^{\mathsf{F}} \mathbf{d}_{k}^{(\ell-1)}+\notag\\
			&&\mathbf{W}^{\mathsf{F}}\mathbf{W}^{\mathsf{V}}\sum_{i=1}^{N_{\mathsf{token}}}  \xi\Big(\big(\mathbf{W}^{\mathsf{Q}} \mathbf{d}_k^{(\ell-1)}\big)^{\mathsf{T}} \big(\mathbf{W}^{\mathsf{K}} \mathbf{d}_i^{(\ell-1)}\big)\Big) \mathbf{d}_i^{(\ell-1)}\Bigg)\notag\\
			\!\!&\!\!\triangleq\!\!&\!\! f_{\mathsf{T}}\left( \mathbf{d}_{k}^{(\ell-1)}+ \sum_{i=1}^{N_{\mathsf{token}}} q_{\mathsf{T}}(\mathbf{d}_{k}^{(\ell-1)},\mathbf{d}_{i}^{(\ell-1)})\right),
		\end{eqnarray}
		where $q_\mathsf{T}(\cdot)$ and $f_\mathsf{T}(\cdot)$ are respectively referred to as the ``processor'' and ``combiner'' of the Transformer.
		%We can see that $q_\mathsf{T}(\cdot)$ uses the attention score for processing.

		\subsubsection{Tokenization} \label{sec:transformer-precoding}
		Since the input of the Transformer should be the representation vectors of tokens while the input of a DNN for precoding is the channel matrix $\mathbf{H}$, we can either set the rows or the columns of $\mathbf{H}$ as the token representations.
		
		\emph{\bf 1D-Transformer:}
		If the tokens are the $K$ users, $N_{\mathsf{token}}=K$, then the columns of $\mathbf{H}$ are the token representations. For the $k$-th token, $\mathbf{d}_k^{(0)}= [\mathsf{Re}(\mathbf{h}_k^{\mathsf{T}}),\mathsf{Im}(\mathbf{h}_k^{\mathsf{T}})]^{\mathsf{T}}$, $\mathbf{h}_k \triangleq [h_{1k},\cdots,h_{Nk}]^{\mathsf{T}}$, while the output of the attention sub-layer in the first layer can be obtained from \eqref{eq:1d-attention} as,
		\begin{equation}\label{eq:1st-layer-att}
			\mathbf{c}_{k}^{(1)}
			=\sum_{i=1}^{K}  \xi\Big(\big(\mathbf{W}^{\mathsf{Q}} \mathbf{h}_k\big)^{\mathsf{T}} \big(\mathbf{W}^{\mathsf{K}} \mathbf{h}_i\big)\Big)\big(\mathbf{W}^{\mathsf{V}} \mathbf{h}_k \big),
		\end{equation}
		where $\mathbf{W}^{\mathsf{Q}}, \mathbf{W}^{\mathsf{K}} \in \mathbb{R}^{N_p \times NJ^{(\ell-1)}}, \mathbf{W}^{\mathsf{V}} \in \mathbb{R}^{NJ^{(\ell-1)} \times NJ^{(\ell-1)}}$, and $N_p$ is the dimension of projection.
		
		The attention score becomes
		\begin{equation}\label{eq:1d-attention-1D}
			\alpha_{ki}\triangleq \xi\Big(\big(\mathbf{W}^{\mathsf{Q}} \mathbf{h}_k\big)^{\mathsf{T}} \big(\mathbf{W}^{\mathsf{K}} \mathbf{h}_i\big)\Big),
		\end{equation}
		which can reflect the MUI between the $k$-th and the $i$-th users.
		
		It has been proved in \cite{mehrabian2024joint,Transformer_PE} that the encoder-only Transformer is equivariant to the permutations of tokens, although not mentioning that the positional encoding should be omitted.
		
		\begin{proposition}\label{prop:transformer-pe}
			When each token is a user, the tailored Transformer satisfies the 1D-PE property in \eqref{1d-pe-UE} but does not satisfy the 2D-PE property.
			\begin{IEEEproof}
				{See Appendix \ref{proof:1D-PE}.}
			\end{IEEEproof}
		\end{proposition}
		
		Intuitively, this is because all the trainable weight matrices are shared among users but not shared among antennas, noticing that their sizes are independent of $K$ but depend on $N$.
		As a result, the tailored Transformer does not satisfy the PE property of the antenna dimension in \eqref{1d-pe-AN}.
		
\begin{figure}[!htbp]
			\centering
			\subfigure[1D-Transformer]
			{\includegraphics[width=0.40\linewidth]{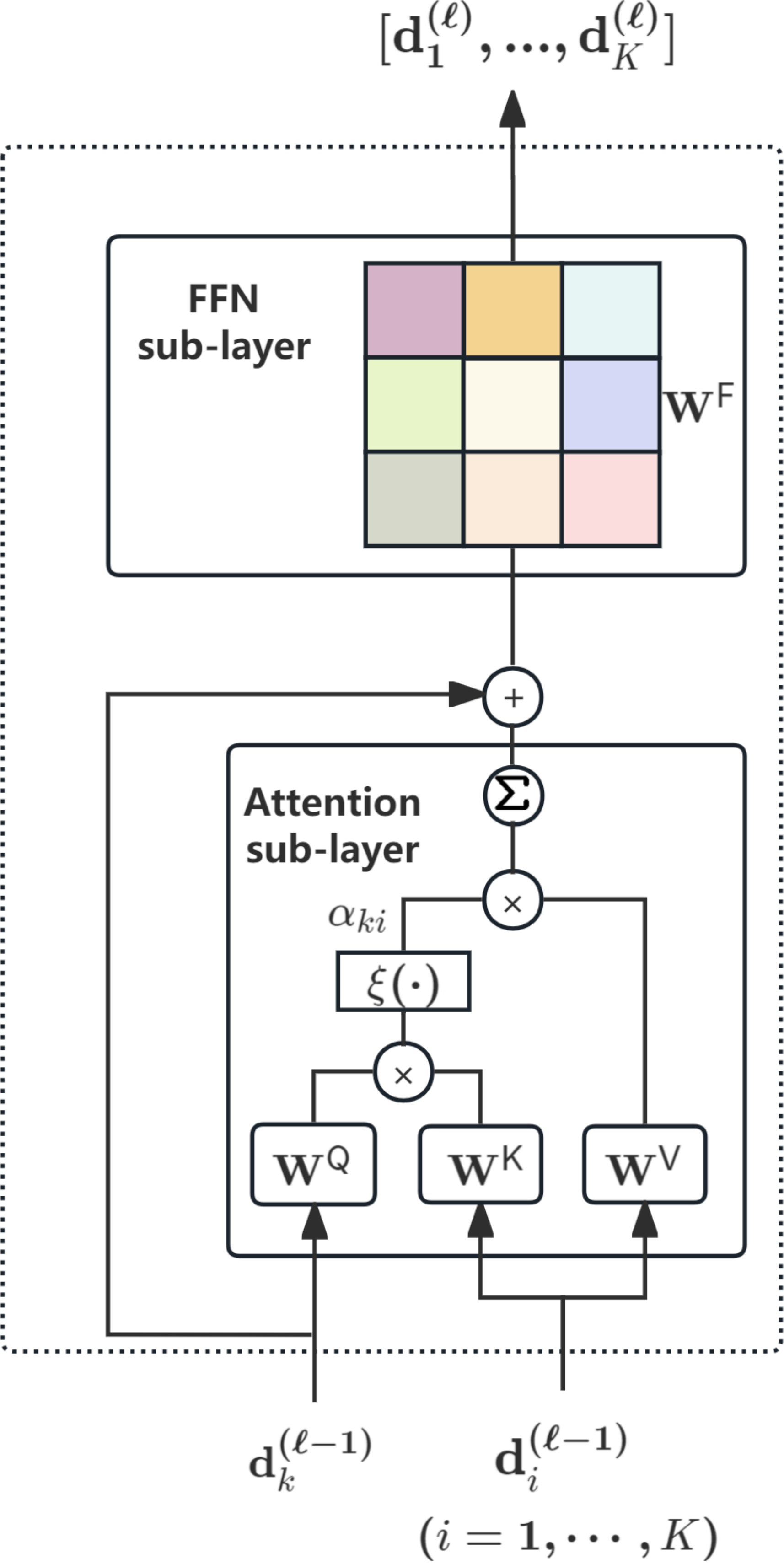}\label{Transformer layer}}
			\subfigure[2D-Gformer]{
				\includegraphics[width=0.40\linewidth]{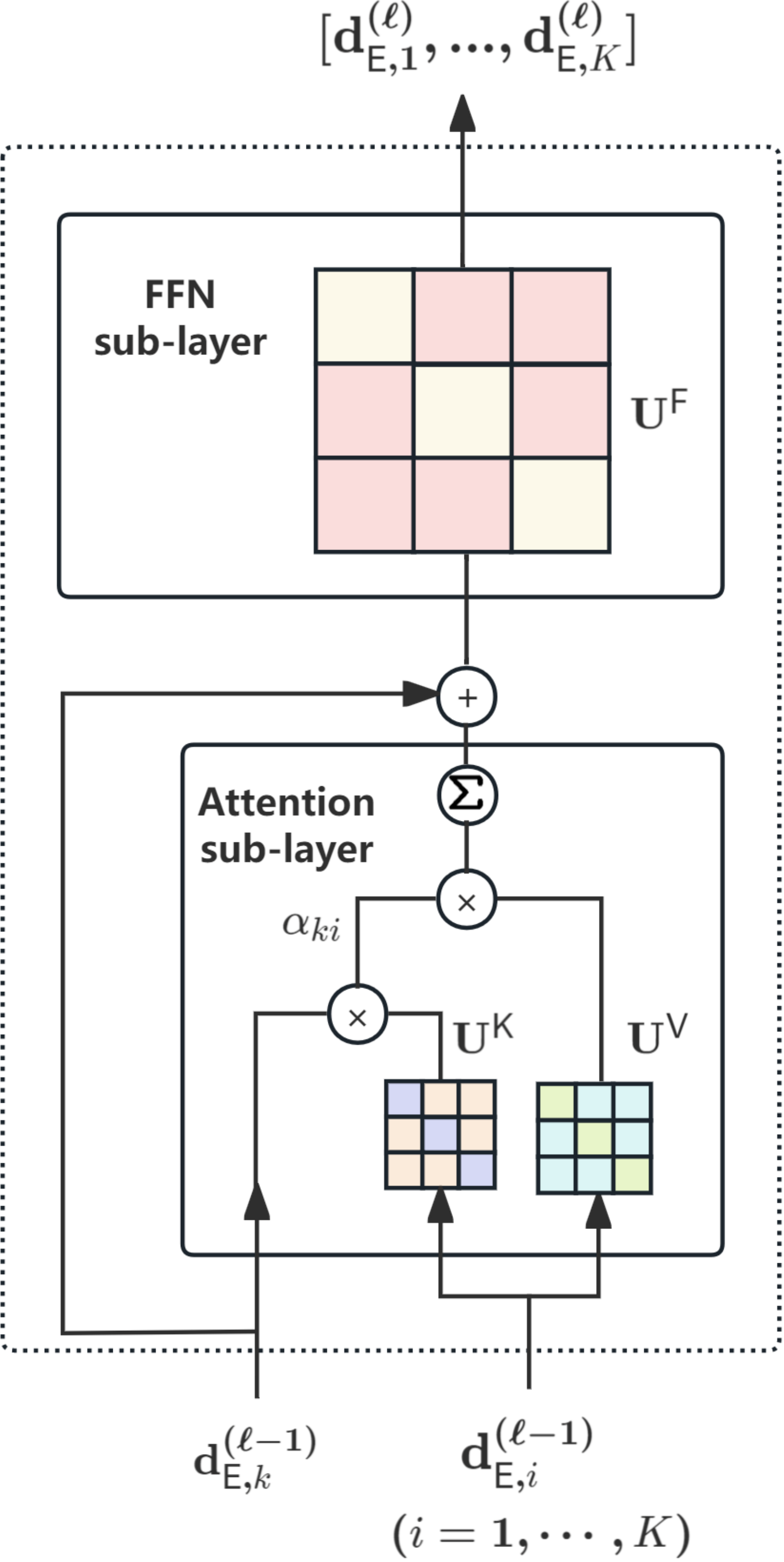}\label{2D-Gformer layer}}\vspace{-2mm}
			\caption{Architectures of the $\ell$-th layer in the 1D-Transformer and the 2D-Gformer. Identical matrix blocks are represented using the same color.}
		\end{figure}

		Such a Transformer is referred to as the 1D-Transformer, whose architecture in the $\ell$-th layer is shown in Fig. \ref{Transformer layer}.
		
		\vspace{-4mm}\vspace{-2mm}
		
		\emph{\bf 1D-Transformer-AN:}
		If the tokens are the $N$ antennas, $N_{\mathsf{token}}=N$, then the rows of the channel matrix are the token representations.  For the $n$-th token,  $\mathbf{d}_n^{(0)}= [\mathsf{Re}(\mathbf{h}^{\mathsf{row}{\,\mathsf{T}}}_n),\mathsf{Im}(\mathbf{h}^{\mathsf{row}\,{\mathsf{T}}}_n)]^{\mathsf{T}}, \mathbf{h}^{\mathsf{row}}_n \triangleq [h_{n1},\cdots,h_{nK}]^\mathsf{T}$, while the output of the attention sub-layer in the first layer is,
		\begin{equation}\label{eq:1st-layer-att-row}
			\mathbf{c}_{n}^{(1)}
			=\sum_{i=1}^{N}  \xi\Big(\big(\mathbf{W}^{\mathsf{Q}} \mathbf{h}^{\mathsf{row}}_n\big)^{\mathsf{T}} \big(\mathbf{W}^{\mathsf{K}} \mathbf{h}^{\mathsf{row}}_i\big)\Big)\big(\mathbf{W}^{\mathsf{V}} \mathbf{h}^{\mathsf{row}}_i \big),
		\end{equation}
		where $\mathbf{W}^{\mathsf{Q}}, \mathbf{W}^{\mathsf{K}} \in \mathbb{R}^{N_p \times KJ^{(\ell-1)}},\mathbf{W}^{\mathsf{V}} \in \mathbb{R}^{KJ^{(\ell-1)} \times KJ^{(\ell-1)}}$. The attention score is $\alpha_{ni}\triangleq \xi\Big(\big(\mathbf{W}^{\mathsf{Q}} \mathbf{h}^{\mathsf{row}}_n\big)^{\mathsf{T}} \big(\mathbf{W}^{\mathsf{K}} \mathbf{h}^{\mathsf{row}}_i\big)\Big)$, which can reflect the correlation between the channels from the $n$-th antenna and the $i$-th antenna to all the users but cannot reflect the MUI.
		
		When each token is an antenna, the tailored Transformer is referred to as 1D-Transformer-AN. Using similar derivations as in Appendix \ref{proof:1D-PE}, we can prove that the 1D-Transformer-AN satisfies the 1D-PE property in \eqref{1d-pe-AN} but not the 2D-PE property.
		
		The 1D-Transformer-AN is with similar architecture in Fig. \ref{Transformer layer}, but the subscript $K$ should be replaced by $N$.
		
		In what follows, we take 1D-Transformer as an example to show the relation of the tailored Transformers with GNNs.
		
		%Since reflecting the MUI is essential for the size-generalizability to users, the 1D-Transformer should be adopted for precoding in order to adapt to the number of users.
		%Then, the representations of the $k$-th token in the first and last layer are the channel vector and the precoding vector of the $k$-th user, respectively.
		
		\vspace{-2mm}
		\subsection{Relation of the 1D-Transformer and GNNs for Precoding}\label{sec:Relation}
		Previous works have noticed that a Transformer is equivalent to a homo-GNN  \cite{gradient, GNNandTrans,transformerisGNN2} by regarding each token as a vertex in the graph, though not mentioning the encoder-only architecture without positional encoding.
		
\begin{figure}[htpb]
			\centering
			\includegraphics[width=0.65\linewidth]{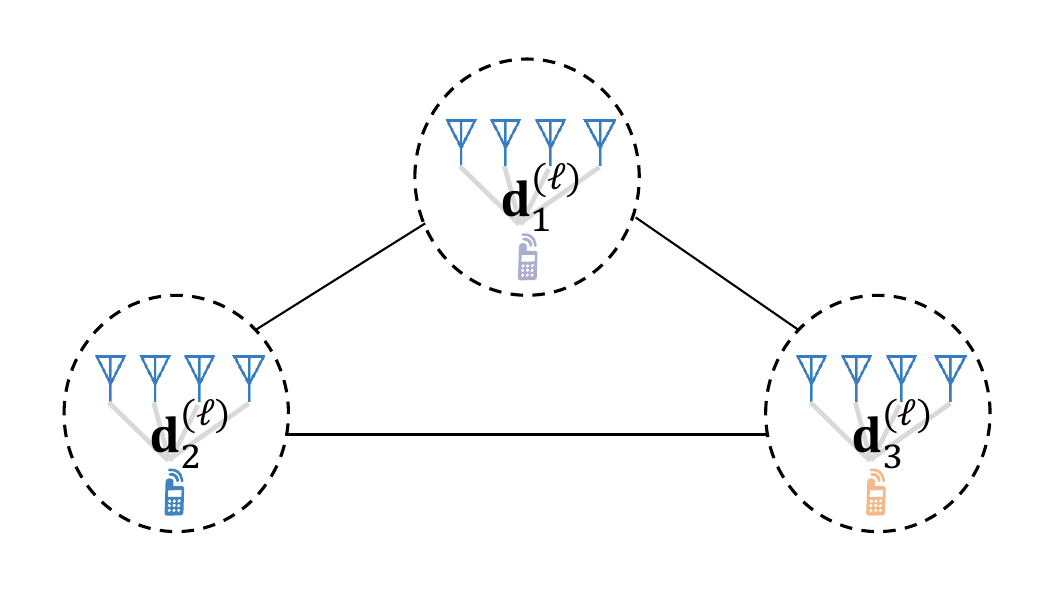}\vspace{-4mm}
			\caption{Illustration of a homogeneous graph modeled for precoding, where each user is defined as a vertex, the feature and action of the $k$-th vertex are respectively $\mathbf{h}_k$ and $\mathbf{v}_k$. $K=3$.}
			\label{homograph}
		\end{figure}

		Similarly, the 1D-Transformer for precoding can be seen as a Vertex-GNN that learns over the graph illustrated in Fig. \ref{homograph}, where each token (i.e., each user) corresponds to a vertex.
		The representation of each token is updated using \eqref{eq:upd-transformer}, where $q_\mathsf{T}(\cdot)$ and $f_\mathsf{T}(\cdot)$ respectively correspond to the processor and combiner of the GNN, whose pooling function is summation.
		
		However, the 1D-Transformer is not equivalent to the Edge-GNNs in section \ref{sec:gnns}. Two major differences and their impacts are summarized below.
		\vspace{-2mm}\vspace{-3mm}
		
		\subsubsection{PE property and its impact}
		The PE property satisfied by a GNN depends on the graph it learns over as well as the parameter sharing of the GNN \cite{liu2023multidimensional}.
		For the Edge-GNNs that learn over the heterogeneous graph illustrated in Fig. \ref{Graph}, the 2D-PE property in \eqref{2d-pe} is satisfied. For the 1D-Transformer, only the 1D-PE property in \eqref{1d-pe-UE}  is satisfied.
		
		%Since satisfying the PE property of a set is a necessary condition for a DNN being generalizable to the size of the set \cite{guo2024recursive},
		As a consequence, the 1D-Transformer has the potential to be generalized to the number of users but is unable to be generalized to the number of antennas.
		Furthermore, as the 1D-Transformer can only satisfy the 1D-PE property while the Edge-GNNs exhibit matched PE property to the precoding policy, the 1D-Transformer has higher training complexity than the Model-GNN and RGNN that are also with attention mechanism, as to be validated via simulations later.
		
		\subsubsection {Reflecting the MUI and its impact} \label{subsec:compare}
		
		\begin{itemize}
			\item 1D-Transformer vs. Edge-GCN: The processor in the Transformer can reflect the MUI, while the processor in the Edge-GCN cannot.
			\item 1D-Transformer vs. RGNN: Both of them can reflect the MUI but in different ways. In the 1D-Transformer, the MUI is reflected by the attention score, which is the inner product of Query and Key. In the RGNN, the MUI is learned by the parameterized function $q_{\mathsf{R}}(\cdot)$.
			\item 1D-Transformer vs. Model-GNN: Both of them reflect the MUI by inner products. In the 1D-Transformer, the inner product of $\mathbf{W}^{\mathsf{Q}}\mathbf{d}_k^{(\ell-1)}$ and $\mathbf{W}^{\mathsf{K}}\mathbf{d}_i^{(\ell-1)}$ are first computed, which is then passed through the softmax function $\xi(\cdot)$. In the Model-GNN, the inner product of $\mathbf{d}_{\mathsf{E},k}^{(\ell-1)}$ and $\mathbf{h}_i$ are computed without  $\xi(\cdot)$ and any trainable parameters.
		\end{itemize}
		
		Since the 1D-Transformer is permutable in the user dimension and is able to reflect the MUI as the RGNN and Model-GNN, the 1D-Transformer can also be generalized to the number of users.
		
		%\begin{remark} The Model-GNN and RGNN are actually \emph{Graph Transformers}, which can be viewed as a kind of Transformers able to cope with graphs or a new architecture incorporating GNN and Transformer \cite{GraphTransformer2024}. Specifically, the Model-GNN is a Graph Transformer where the TGNN sub-layer is with an attention mechanism and the Edge-GCN sub-layer updates over a graph. RGNN can be viewed as a ``recursive" Graph Transformer, where the attention operation in each layer is realized by another Transformer in the second recursion, and both recursions in each layer are updated over a graph.
		%\end{remark}
		
		\begin{remark}
			In \cite{GAT_MISO_EE,GAT_Hybrid}, GATs were adopted to learn precoding policies over the homogeneous graph illustrated in Fig. \ref{homograph}, which are a kind of Vertex-GNNs. The GAT employs a similar attention mechanism to the 1D-Transformer, but the definition of attention scores differs. Specifically, the representation of each vertex is updated as follows,
			\begin{equation}
				\begin{aligned}
					\mathbf{d}_k^{(\ell)} &= \sum_{i=1}^{K} \underbrace{\sigma\bigg(\mathbf{W}^{\mathsf{Q}} \mathbf{d}_k^{(\ell-1)}+\mathbf{W}^{\mathsf{K}}
						\mathbf{d}_i^{(\ell-1)}\bigg)}_{\alpha_{ki}^{\mathsf{GAT}}} \mathbf{W}^{\mathsf{V}} \mathbf{d}_i^{(\ell-1)},\label{GAT_update}
				\end{aligned}
			\end{equation}
			where the attention score $\alpha_{ki}^{\mathsf{GAT}}$ is in the form of weighted summation, while the attention score in the Transformer is the form of the inner product. Hence, the GATs can be seen as variants of the 1D-Transformer without the FFN sub-layer in each layer. Since the GATs only satisfy the 1D-PE property in \eqref{1d-pe-UE}, they cannot be generalized to the number of antennas.
		\end{remark}
		
		\vspace{-4mm}\section{Design of Graph Transformers for Precoding}\label{sec:2D-Gformer}
		In this section, we show how to design Graph Transformers to learn precoding policies with matched PE properties, inspired by the relation analyzed in section \ref{sec:Relation}.
		
		We begin with the design of a Graph Transformer, called 2D-Gformer, for learning the baseband precoding, which learns over the heterogeneous graph in Fig. \ref{Graph} and satisfies the 2D-PE property. We proceed to extend the design for three-dimensional (3D)-Gformer satisfying the PE properties of three dimensions by taking learning hybrid analog and baseband precoding as an example. Both Gformers are designed from the encoder-only Transformer without positional encoding, each consists of $L$ layers.
		
		\vspace{-2mm}\subsection{2D-Gformer for Learning Baseband Precoding}
		For the 2D-Gformer, each layer is composed of an attention sub-layer and an FFN sub-layer, and its input and output are respectively $\mathbf{H}$ and $\mathbf{V}$. We start by presenting an F-2D-Gformer by extending the 1D-Transformer and then propose a 2D-Gformer by removing unnecessary components.
		
		\subsubsection{Tokenization} Since all the features and actions in Fig. \ref{Graph} are defined on edges, one can define each edge as a token. However, in this way, the attention score reflects the dependence of edge representations but cannot reflect the MUI, incurring the non-generalizability to the number of users.
		
		For the heterogeneous graph with two types of vertices, the way to define tokens is not unique. If we define each antenna vertex as a token, then the MUI cannot be reflected, similar to the 1D-Transformer-AN.
		Same as the 1D-Transformer, we define each token as a user vertex and hence the representation of the $k$-th token, denoted as $\mathbf{d}_{\mathsf{E},k}^{(0)}=[\mathbf{d}_{\mathsf{E},1k}^{(0)\mathsf{T}},\cdots, \mathbf{d}_{\mathsf{E},Nk}^{(0)\mathsf{T}}]^{\mathsf{T}}=[\mathsf{Re}(\mathbf{h}_k^{\mathsf{T}}),\mathsf{Im}(\mathbf{h}_k^{\mathsf{T}})]^{\mathsf{T}}$, consists of the features of all the edges connected to the $k$-th user vertex. Then, the MUI can be reflected by the attention score.
		The corresponding output is $\mathbf{d}_{\mathsf{E},k}^{(L)}=[\mathsf{Re}(\mathbf{v}_k^{\mathsf{T}}),\mathsf{Im}(\mathbf{v}_k^{\mathsf{T}})]^{\mathsf{T}}$, which is the actions of all the edges connected to the $k$-th user vertex.

		\subsubsection{The F-2D-Gformer}
		Denote the output of the \emph{attention sub-layer} in the $\ell$-th layer of the F-2D-Gformer as $\mathbf{c}_{\mathsf{E},k}^{(\ell)}$,
		which has the same form as in \eqref{eq:1d-attention}, but $\mathbf{d}_k^{(\ell-1)}$ and $\mathbf{c}_k^{(\ell)}$ are replaced by $\mathbf{d}_{\mathsf{E},k}^{(\ell-1)}$ and $\mathbf{c}_{\mathsf{E},k}^{(\ell)}$.
		
		Similarly, the output of the \emph{FFN sub-layer} has the same form as in \eqref{eq:ffn},  but $\mathbf{d}_k^{(\ell-1)}$, $\mathbf{d}_k^{(\ell)}$ and $\mathbf{c}_k^{(\ell)}$ in \eqref{eq:ffn} are replaced by $\mathbf{d}_{\mathsf{E},k}^{(\ell-1)}$, $\mathbf{d}_{\mathsf{E},k}^{(\ell)}$, and $\mathbf{c}_{\mathsf{E},k}^{(\ell)}$.
		
		%\subsubsection{Satisfying the 2D-PE property}
		Then, the \emph{update equation} for the $k$-th token in the $\ell$-th layer of the F-2D-Gformer is,
		\begin{equation}
			\label{eq:2d-trans}
			\begin{aligned}
				\mathbf{d}_{\mathsf{E},k}^{(\ell)}& = \sigma \bigg(\mathbf{U}^{\mathsf{F}}\mathbf{d}_{\mathsf{E},k}^{(\ell-1)}+\\&\mathbf{U}^{\mathsf{F}}\mathbf{U}^{\mathsf{V}}\sum_{i=1}^{K} \xi\Big(    \big(\mathbf{U}^{\mathsf{Q}}\mathbf{d}_{\mathsf{E},k}^{(\ell-1)}\big)^\mathsf{T}
				\big(\mathbf{U}^{\mathsf{K}} \mathbf{d}_{\mathsf{E},i}^{(\ell-1)}\big)\Big) \mathbf{d}_{\mathsf{E},i}^{(\ell-1)}\bigg)\\&
				\triangleq F\big(\mathbf{d}_{\mathsf{E},1}^{(\ell-1)},\cdots,\mathbf{d}_{\mathsf{E},K}^{(\ell-1)}\big),
			\end{aligned}
		\end{equation}
		where $ \mathbf{U}^{\mathsf{F}} \in \mathbb{R}^{NJ^{(\ell)} \times NJ^{(\ell-1)}}$, $\mathbf{U}^{\mathsf{Q}},\mathbf{U}^{\mathsf{K}} \in \mathbb{R}^{N_p \times NJ^{(\ell-1)}}$, and $\mathbf{U}^{\mathsf{V}} \in \mathbb{R}^{NJ^{(\ell-1)} \times NJ^{(\ell-1)}}$ are trainable weight matrices.  All these matrices are shared among user vertices.
		
		The differences from the 1D-Transformer lie in the updated token representation, the structure of the weight matrices, and the dimension of projection $N_p$, all for satisfying the 2D-PE property as explained in the following.
		
		In the 1D-Transformer, the representations of user vertices (e.g., $\mathbf{d}_k^{(\ell-1)}$ for the $k$-th user vertex) are updated. By contrast, the representations of the edges connected to user vertices (e.g., $\mathbf{d}_{\mathsf{E},k}^{(\ell)}$ for all edges connected to the $k$-th user vertex) are updated in the F-2D-Gformer.
		
		Thanks to the relation between the 1D-Transformer and the RGNN, we can borrow the idea of designing the RGNN to satisfy the PE property along every dimension recursively for satisfying the 2D-PE property. Using similar derivations as in Appendix \ref{proof:1D-PE}, it is not hard to prove that the F-2D-Gformer satisfies the PE property in the user dimension. The following proposition indicates how to design the F-2D-Gformer to further satisfy the PE property in the antenna dimension.
		
		\begin{proposition}\label{lemma:1-revised}
			If $\mathbf{\Pi}_{\mathsf{AN}}\mathbf{U}^{\mathsf{Q}}=\mathbf{U}^{\mathsf{Q}}\mathbf{\Pi}_{\mathsf{AN}}$, $\mathbf{\Pi}_{\mathsf{AN}}\mathbf{U}^{\mathsf{K}}=\mathbf{U}^{\mathsf{K}}\mathbf{\Pi}_{\mathsf{AN}}$, $\mathbf{\Pi}_{\mathsf{AN}}\mathbf{U}^{\mathsf{V}}=\mathbf{U}^{\mathsf{V}}\mathbf{\Pi}_{\mathsf{AN}}$,
			and    $\mathbf{\Pi}_{\mathsf{AN}}\mathbf{U}^{\mathsf{F}}=\mathbf{U}^{\mathsf{F}}\mathbf{\Pi}_{\mathsf{AN}}$,
			the F-2D-Gformer is equivariant to the permutations of antennas, i.e.,
			\begin{eqnarray}\label{eq:pi-upd-2D-Gformer-revised}
				\mathbf{\Pi}_{\mathsf{AN}}\mathbf{d}_{\mathsf{E},k}^{(\ell)} = F\big(\mathbf{\Pi}_{\mathsf{AN}}\mathbf{d}_{\mathsf{E},1}^{(\ell-1)},\cdots,\mathbf{\Pi}_{\mathsf{AN}}\mathbf{d}_{\mathsf{E},K}^{(\ell-1)}\big).
			\end{eqnarray}
			\begin{IEEEproof}
				See Appendix \ref{proof:lemma-revised}.
			\end{IEEEproof}
		\end{proposition}
		When $\mathbf{U}^{\mathsf{Q}}$, $\mathbf{U}^{\mathsf{K}}$, $\mathbf{U}^{\mathsf{V}}$, and $\mathbf{U}^{\mathsf{F}}$ are designed with the parameter-sharing structure in \eqref{eq:matrix structure}, the four conditions in the proposition can be satisfied \cite{Deepsets}. For satisfying the first two conditions, the projection dimension should be $N_p = NJ^{(\ell-1)}$, otherwise $\mathbf{U}^{\mathsf{Q}}$ and $\mathbf{U}^{\mathsf{K}}$ cannot exhibit the structure.
		
		\begin{remark}
			The attention mechanism can be extended to a multi-head version. Denote the number of heads as $M_h$. In the $m$-th head, the input $\mathbf{d}_{\mathsf{E},k}^{(\ell-1)}$ is projected to the Query, Key, and Value respectively with the weight matrices $\mathbf{U}^{\mathsf{Q}}_m$, $\mathbf{U}^{\mathsf{K}}_m$, $\mathbf{U}^{\mathsf{V}}_m \in \mathbb{R}^{N \times NJ^{(\ell-1)}}$. The output of the attention sub-layer in the $m$-th head is as follows,
			\begin{equation}
				\mathbf{c}_{{\mathsf{E},k},m}^{(\ell)} =\sum_{i=1}^{K}  \xi\Big(\big(\mathbf{U}^{\mathsf{Q}}_m \mathbf{d}_{\mathsf{E},k}^{(\ell-1)}\big)^{\mathsf{T}}
				\big(\mathbf{U}^{\mathsf{K}}_m \mathbf{d}_{\mathsf{E},i}^{(\ell-1)}\big)\Big) (\mathbf{U}^{\mathsf{V}}_m\mathbf{d}_{\mathsf{E},i}^{(\ell-1)}). \notag
			\end{equation}
			To satisfy the 2D-PE property, $\mathbf{U}^{\mathsf{Q}}_m$, $\mathbf{U}^{\mathsf{K}}_m$, $\mathbf{U}^{\mathsf{V}}_m$ should be with the structure in \eqref{eq:matrix structure}. The outputs of all heads are then concatenated to obtain $\mathbf{c}_{\mathsf{E},k}^{(\ell)} = [\mathbf{c}_{\mathsf{E},k,1}^{(\ell)\mathsf{T}},\cdots,\mathbf{c}_{{\mathsf{E},k},M_h}^{(\ell)\mathsf{T}}]^{\mathsf{T}}$.
			The weight matrices of all the heads can be stacked together as $\mathbf{U}^{\sf Q}=[\mathbf{U}^{\mathsf{Q}}_1,\cdots, \mathbf{U}^{\mathsf{Q}}_{M_h}]$, $\mathbf{U}^{\sf K}=[\mathbf{U}^{\mathsf{K}}_1,\cdots, \mathbf{U}^{\mathsf{K}}_{M_h}]$ and $\mathbf{U}^{\sf V}=[\mathbf{U}^{\mathsf{V}}_1,\cdots, \mathbf{U}^{\mathsf{V}}_{M_h}]$, which are with size of ${ NJ^{(\ell-1)}\times N M_h}$. This indicates that the multiple heads in Transformers play the same role of the representation dimensions of GNNs.
		\end{remark}

		\subsubsection{The proposed 2D-Gformer}\label{sec:simp-2D-Trans}
		The F-2D-Gformer is extended from the 1D-Transformer for learning over the heterogeneous
		graph and satisfying the 2D-PE property. However, are all its components (say $\mathbf{U}^{\mathsf{Q}}$, $\mathbf{U}^{\mathsf{K}}$, $\mathbf{U}^{\mathsf{V}}$, and $\xi(\cdot)$) necessary for learning the precoding policy?

		Because the Model-GNN is designed on the basis of a mathematical model, it is relatively interpretable. We resort to the relation of the 1D-Transformer with the Model-GNN to guide the simplification of the F-2D-Gformer.
		By comparing the update equations in \eqref{Model-GNN update eq} and \eqref{eq:2d-trans}, we can observe that the F-2D-Gformer and Model-GNN are with similar architectures, and the only difference lies in the way of reflecting the MUI. This helps us to analyze the role of each component in the F-2D-Gfomer and decide which components can be removed to reduce computational complexity.
		
		In particular, in the Model-GNN, the inner product of $\mathbf{d}_{\mathsf{E},k}^{(\ell-1)}$ and $\mathbf{h}_i$ is computed to reflect the MUI, while the inner product of $\mathbf{U}^{\mathsf{Q}}\mathbf{d}_{\mathsf{E},k}^{(\ell-1)}$ and $\mathbf{U}^{\mathsf{K}}\mathbf{d}_{\mathsf{E},i}^{(\ell-1)}$ followed by the softmax function $\xi(\cdot)$ is computed in the F-2D-Gformer. This suggests that the trainable weight matrix $\mathbf{U}^{\mathsf{K}}$ in the F-2D-Gformer plays the role of recovering $\mathbf{h}_i$ from $\mathbf{d}_{\mathsf{E},i}^{(\ell-1)}$ for computing the MUI, whereas $\mathbf{U}^{\mathsf{Q}}$ and $\xi(\cdot)$ are useless.
		On the other hand, $\mathbf{U}^{\mathsf{F}}$ and $\mathbf{U}^{\mathsf{F}}\mathbf{U}^{\mathsf{V}}$ in \eqref{eq:2d-trans} are analogous to $\mathbf{W}$ and $\mathbf{U}$ in \eqref{Model-GNN update eq}, which are used to adjust the direction and power of the updated precoding vector for the $k$-th user  \cite{guo2023model}, and hence should be preserved.
		
		After the simplification, the update equation  of the proposed 2D-Gformer can be expressed as,
		\begin{equation}\label{eq:upd-2d-trans-simp}
			\mathbf{d}_{\mathsf{E},k}^{(\ell)} = \sigma \Big(\mathbf{U}^{\mathsf{F}}\mathbf{d}_{\mathsf{E},k}^{(\ell-1)}+\mathbf{U}^{\mathsf{F}}\mathbf{U}^{\mathsf{V}}\sum_{i=1}^{K}  \underbrace{ \mathbf{d}_{\mathsf{E},k}^{(\ell-1){\mathsf{T}}}
				\big(\mathbf{U}^{\mathsf{K}} \mathbf{d}_{\mathsf{E},i}^{(\ell-1)}\big) }_{\alpha_{ki}}\mathbf{d}_{\mathsf{E},i}^{(\ell-1)}\Big),
		\end{equation}
		whose architecture is shown in Fig. \ref{2D-Gformer layer}. It is not hard to prove that the 2D-Gformer still satisfies the 2D-PE property.

		Since the softmax function $\xi(\cdot)$ is computationally expensive in Transformers especially for large-scale problems \cite{softmaxfree}, removing $\xi(\cdot)$ can reduce computational complexity. %Additionally, as we analyzed previously, $\xi(\cdot)$ is not required for learning precoding. Hence, the 2D-former can achieve the same performance as the F-2D-Gformer with lower computational complexity for learning precoding, as to be validated later.
		
		\vspace{-2mm}\subsection{Extension to 3D-Gformer for Learning Hyprid Precoding}
		The 2D-Gformer can be extended to a 3D-Gformer to satisfy the 3D-PE property, by inheriting the spirit of the RGNN.
		We take hybrid precoding in the MU-MISO system as an example to demonstrate how to design a 3D-Gformer.
		
		The hybrid precoding policy can be expressed as the mapping from $\mathbf{H} \in \mathbb{C}^{N\times K}$ to the baseband precoding matrix $\mathbf{V}_{\mathsf{BB}} \in \mathbb{C}^{K \times N_\mathsf{RF}}$ and analog precoding matrix $\mathbf{V}_{\mathsf{RF}} \in \mathbb{C}^{N_\mathsf{RF}\times N}$, where $N_{\mathsf{RF}}$ is the number of radio frequency (RF) chains.
		
		The policy is not affected by the permutations of users, antennas, and RF chains. To satisfy the permutation property, the 3D-Gformer is designed to learn over a heterogeneous graph with user vertices, antenna vertices, and RF chain vertices. Each type of vertices is connected to the other two types of vertices by a hyper-edge.
		
		In this policy, the input matrix $\mathbf{H}$ spans a 2D space with antenna dimension and user dimension, while the output matrices $\mathbf{V}_{\mathsf{BB}}$ and $\mathbf{V}_{\mathsf{RF}}$ span a 3D space with antenna, user, and RF chain dimensions. To avoid information loss during information aggregation, we increase the dimension of the feature for each hyper-edge (say the hyper-edge connecting the $n$-th antenna vertex, $k$-th user vertex, and $r$-th RF chain vertex) by setting $\mathbf{d}^{(0)}_{\mathsf{E},nkr} = [\mathsf{Re}(h_{nk}) + a_r, \mathsf{Im}(h_{nk})]^{\mathsf{T}} \in \mathbb{R}^{J^{(0)}}$ as in \cite{liu2023multidimensional}, where $J^{(0)}=2$, and $\mathbf{a} \triangleq [a_1,\cdots, a_{N_\mathsf{RF}}]$ is a virtue vector with random elements for increasing the dimension.
		
		%optimization problem can be expressed as,
		%\begin{equation}
		%    \begin{aligned}
			% &\max_{\mathbf{V}_{\mathsf{RF}},\mathbf{v}_{\mathsf{BB}}} \\&\sum_{k=1}^{K}\log_{2}\left(1+\frac{|\mathbf{h}_{k}^{\mathsf{H}}\mathbf{V}_{\mathsf{RF}}\mathbf{v}_{\mathsf{BB},k}|^{2}}{\sum_{i=1,i\neq k}^{K}|\mathbf{h}_{k}^{\mathsf{H}}\mathbf{V}_{\mathsf{RF}}\mathbf{v}_{\mathsf{BB},i}|^{2}+\sigma^{2}_0}\right),& \\
			% &\quad \mathrm{s.t.}  \quad \mathsf{Tr}(\mathbf{V}_{\mathsf{RF}}\mathbf{V}_{\mathsf{BB}}\mathbf{V}_{\mathsf{BB}}^\mathsf{H}\mathbf{V}_{\mathsf{RF}}^{\mathsf{H}})=P_{t}, \\
			% &\quad\quad\quad|\left(\mathbf{V}_{\mathsf{RF}}\right)_{j,l}|=1,j=1,\cdots,N,l=1,\cdots,N_{s},
			% \end{aligned}
		% \end{equation}
	% where $\mathbf{V}_{\mathsf{BB}} \triangleq [\mathbf{v}_{\mathsf{BB},1},\cdots,\mathbf{v}_{\mathsf{BB},K}]$ and $\mathbf{V}_{\mathsf{RF}} \triangleq [\mathbf{v}_{\mathsf{RF},1},\cdots,\mathbf{v}_{\mathsf{RF},N}]\triangleq [(\mathbf{V}_{\mathsf{RF}})_{j,l}]$ represents the baseband precoding matrix and analog precoding matrix, $N_s$ is the number of RF chains.
	
	Again, we define each token as a user for reflecting the MUI. If we set the features of all the hyper-edges connected to each user vertex as the representation of each token, then the complexity to compute an attention score will be high and the parameter-sharing of the weight matrices should be re-designed for satisfying the PE properties of the three dimensions. To circumvent this issue, we conceive another way to define the token representation.
	
	Noticing that the MUI depends on the correlation of channel vectors of users and is irrelevant to RF chains, we allow each token, say the $k$-th token, to have $N_\mathsf{RF}$ representations. In particular, the $r$-th representation of the $k$-th token is defined as $\mathbf{d}_{\mathsf{E},kr}^{(0)} = [\mathbf{d}_{\mathsf{E},1kr}^{(0)\mathsf{T}},\cdots,\mathbf{d}_{\mathsf{E},Nkr}^{(0)\mathsf{T}}]^\mathsf{T} \in \mathbb{R}^{NJ^{(0)}}$, which includes the features of all the hyper-edges connected to the $k$-th user vertex and the $r$-th RF chain vertex. All the representations of the $K$ tokens can be expressed as an order-three tensor $\mathbf{D}^{(0)} = [\mathbf{d}^{(0)}_{\mathsf{E},nkr}]\in \mathbb{C}^{NJ^{(0)} \times K \times N_\mathsf{RF}}$, which are updated in the 3D space. By using the same method as in \cite{liu2023multidimensional}, it can be proved that the 3D-Gformer is with matched permutation property of the hybrid precoding policy if the input-output relation of each layer satisfies 3D-PE property.
	
\begin{figure}
		\centering
		\includegraphics[width=1.0\linewidth]{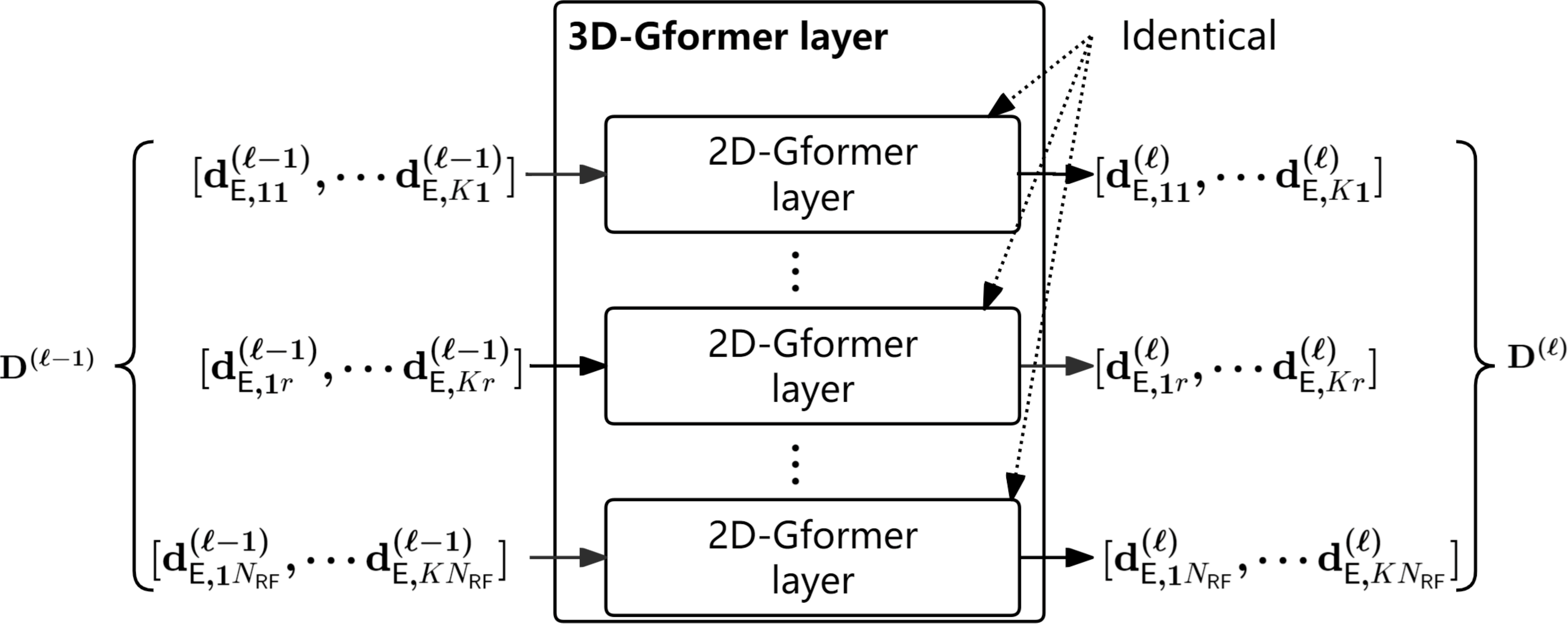}\vspace{-2mm}
		\caption{Architecture of the $\ell$-th layer in the 3D-Gformer. }
		\label{3D-Gformer}
	\end{figure}

	To satisfy the 3D-PE property, the 3D-Gformer can be designed by extending the 2D-Gformer for further satisfying equivariance to the permutation of RF chains. To this end, all the token representations are divided into $N_\mathsf{RF}$ groups, and each group of token representations
	is updated by a 2D-Gformer layer. In other words, each layer of the 3D-Gformer is composed of $N_\mathsf{RF}$ shared 2D-Gformer layers, whose architecture is shown in Fig. \ref{3D-Gformer}.
	Specifically, the $r$-th representation of the $k$-th token is updated in the $\ell$-th layer as follows,
	\begin{equation}\label{eq:upd-3d-transformer}
		\begin{aligned}
			\mathbf{d}_{\mathsf{E},kr}^{(\ell)} = \sigma \bigg(&\mathbf{U}^{\mathsf{F}}\mathbf{d}_{\mathsf{E},kr}^{(\ell-1)}+\\& \mathbf{U}^{\mathsf{F}}\mathbf{U}^{\mathsf{V}}\sum_{i=1}^{K} \mathbf{d}_{\mathsf{E},kr}^{(\ell-1)\mathsf{T}}
			\big(\mathbf{U}^{\mathsf{K}} \mathbf{d}_{\mathsf{E},ir}^{(\ell-1)}\big) \mathbf{d}_{\mathsf{E},ir}^{(\ell-1)}\bigg),
		\end{aligned}
	\end{equation}
	where $\mathbf{U}^{\mathsf{Q}}$, $\mathbf{U}^{\mathsf{K}}$, $\mathbf{U}^{\mathsf{V}}$, and $\mathbf{U}^{\mathsf{F}}$ are designed with the same structure as in \eqref{eq:matrix structure}, and $r=1,\cdots,N_\mathsf{RF}$.
	
	%The differences from the 2D-Gformer lie in the updated token representation.
	
	Using similar derivations as in Appendix \ref{proof:1D-PE}, it is not hard to prove that the 3D-Gformer is indeed further permutation equivariant to the RF chains since the same 2D-Gformer layer is used to update the $N_\mathsf{RF}$ representations for each token.
	
	%Different from the 2D-Gformer where the representations of the edges connected to user vertices (e.g., $\mathbf{d}_{\mathsf{E},k}^{(\ell)}$ for all edges connected to the $k$-th user vertex) are updated, the representations of the hyper-edges connected to user vertices (e.g., $\mathbf{d}_{\mathsf{E},kr}^{(\ell)}$ for all hyper-edges connected to the $k$-th user vertex) are updated in the 3D-Gformer.

	The representations are updated to $\mathbf{D}^{(L)}$ through $L$ layers. Since $\mathbf{D}^{(L)}$ is an order-three tensor while $\mathbf{V}_\mathsf{BB}$ and $\mathbf{V}_\mathsf{RF}$ are matrices, an output layer is needed to map  $\mathbf{D}^{(L)}$ to $\mathbf{V}_\mathsf{BB}$ and $\mathbf{V}_\mathsf{RF}$. In particular, $\mathbf{V}_{\mathsf{RF}}$ can be obtained by first summing over the user dimension in $\mathbf{D}^{(L)}$ as $[\mathsf{Re}(\mathbf{V}_{\mathsf{RF}})_{rn}, \mathsf{Im}(\mathbf{V}_{\mathsf{RF}})_{rn}] = \sum_{k=1}^{K} \mathbf{d}_{\mathsf{E},nkr}^{(L)}$ and then normalized to satisfy the constant modulus constraint. Similarly, $\mathbf{V}_{\mathsf{BB}}$ can be obtained by first summing over the antenna dimension as
	$[\mathsf{Re}(\mathbf{V}_{\mathsf{BB}})_{kr},\mathsf{Im}(\mathbf{V}_{\mathsf{BB}})_{kr}] = \sum_{n=1}^{N} \mathbf{d}_{\mathsf{E},nkr}^{(L)}$ and then normalized to satisfy the total power constraint.

	\vspace{-2mm}
	
	%\begin{figure}[!t]	\label{fig:XD_Transformer}
	%	\centering
	%	\subfigure[3D-Gformer]{\includegraphics[width=1.0\linewidth]{Figures/3D_Transformer.pdf}%
		%		\label{fig_first_case-3D}}
	%
	%	\subfigure[$X$D-Gformer, $X>2$]{\includegraphics[width=0.65\linewidth]{Figures/XD-Transformer.pdf}%
		%		\label{fig_second_case-XD}}
	%	\caption{Architectures of the 3D-Gformer and $X$D-Gformer.}
	%
	%\end{figure}

	\section{Simulation Results}\label{sec:simulation}
	In this section, we compare the learning performance, inference and training complexity, and size-generalizability of the GNNs, Transformers, and Graph Transformers.
	
	We consider the SE-maximization problem under the constraint of total power $P_t$. The output of each DNN is normalized as $\mathbf{v}_k' = \mathbf{v}_{k}^{'}\sqrt{P_t/\sum_{i=1}^{K} \mathbf{v}_{i}^{\mathsf{H}}\mathbf{v}_{i}}$ for baseband precoding to satisfy the power constraint and $(\mathbf{V}_{\mathsf{RF}}')_{ij} = (\mathbf{V}_{\mathsf{RF}})_{ij} / |(\mathbf{V}_{\mathsf{RF}})_{ij}|$, $\mathbf{V}_{\mathsf{BB}} = \mathbf{V}_{\mathsf{BB}}\sqrt{P_t/ \mathsf{Tr}(\mathbf{V}_{\mathsf{BB}}^{\mathsf{H}}\mathbf{V}_{\mathsf{RF}}^\mathsf{'H}\mathbf{V}_{\mathsf{RF}}'\mathbf{V}_{\mathsf{BB}})}$ for hybrid precoding to satisfy both the power and constant modulus constraints.
	The learning performance is measured by the SE ratio, which is the ratio of the SE achieved by a DNN to the SE achieved by the weighted minimal mean square error algorithm in \cite{wmmse} for baseband precoding or the manifold optimization algorithm in \cite{MO} for hybrid precoding.

	In particular, we compare the proposed 2D-Gformer and 3D-Gformer with the Edge-GCN, RGNN, and Model-GNN in section \ref{sec:gnns}, the 1D-Transformer and 1D-Transformer-AN in section \ref{1D-Transformers}, as well as the following DNNs.
	\begin{itemize}
		\item \textbf{Vanilla-Transformer}: This is the Transformer proposed in \cite{vaswani2017attention} with encoder-decoder architecture and positional encoding. For learning precoding, each token representation is set as the channel vector of each user.
		\item  \textbf{Indirect-GCN}: This is another model-driven GNN, where an Edge-GCN is used to learn the power allocation policy $(\mathbf{p}^*,\boldsymbol{\lambda}^*)=g_p(\mathbf{H})$ with matched permutation property  \cite{ZBCGC2023}, with which the precoding matrix is recovered using \eqref{eq_duality}.
	\end{itemize}
	
	\vspace{-2mm}
	\subsection{Simulation Setup and Hyper-parameters}
	For learning baseband precoding, all samples are generated from uncorrelated Rayleigh fading channels, i.e., each element in $\mathbf{H}$ follows the complex Gaussian distribution $\mathcal{CN}(0,1)$, $N=16, K=8$. For hybrid precoding, all samples are generated from the Saleh-Valenzuela channel model, where the numbers of clusters and rays are set as 4 and 5, respectively, $N=16, N_{\mathsf{RF}}=8, K=3$. The signal-to-noise ratio (SNR) is 10 dB. 40,000 samples are used for training, and 2,000 samples are used for testing.
	
	Unless otherwise specified, this setup is used for all the following simulations.
	
	All the DNNs are trained in an unsupervised manner with an Adam optimizer, where the loss function is the negative SE averaged over all the training samples. The activation function in the hidden layers is set as $\mathrm{tanh}(x) = \frac{e^x - e^{-x}}{e^x + e^{-x}}$, whose outputs are between $[-1, 1]$. For the Indirect-GCN, the activation function in the output layer is set as $\mathrm{Relu}(x) = \mathrm{max}(x,0)$ to ensure the learned power allocated to each user is non-negative, while no activation function is used in the output layer of other DNNs.
	After fine-tuning, the number of heads is set as $M_h=32$  for all Transformers, and other hyper-parameters are listed in Table \ref{HYPERPARAMS}.
	
	% The dimension of $\mathbf{d}_{\mathsf{E},k}^{(\ell)}$ or $\mathbf{d}_{k}^{(\ell)}$ is $2N$ in the input and output layers, and is the hyper-parameters in hidden layers.
	
	All the simulations are conducted on a computer with one Intel i9-9920 CPU and one Nvidia RTX 2080Ti GPU.
	\begin{table}[htbp]
		\centering
		\begin{threeparttable}
			\caption{Fine-tuned hyper-parameters of the DNNs}\label{HYPERPARAMS}
			\begin{tabular}{c|c|c|c}
				\hline
				\hline
				DNNs & Layers &   \makecell{$J^{(\ell)}$ \\ in hidden layers} & \makecell{Learning\\ rate} \\
				\hline
				Edge-GCN & 4     & [128,128,128,128] & 0.002\\
				\hline
				Model-GNN & 3     & [32,32,32] & 0.002  \\
				\hline
				RGNN  & 3     & [32,32,32] & 0.002  \\
				\hline
				Indirect-GCN & 3 & [32,32,32] & 0.002 \\
				\hline
				
				Vanilla-Transformer &6 & \makecell{Encoder: [32,32,32] \\ Decoder: [32,32,32]}& 0.002\\
				\hline
				1D-Transformer & 3     & [32,32,32] & 0.0005  \\
				\hline
				1D-Transformer-AN & 3     & [32,32,32] & 0.002  \\
				\hline
				2D-Gformer & 3     & [32,32,32] & 0.002 \\
				\hline
				3D-Gformer &4 &[128,128,128,128] & 0.005\\
				\hline
				
				\hline
				\hline
			\end{tabular}%
			
		\end{threeparttable}
	\end{table}\vspace{-0.01mm}
	% \begin{table}[htbp]
		%   \centering
		%   \begin{threeparttable}
			%   \caption{Fine-tuned hyper-parameters of the DNNs}\label{HYPERPARAMS}
			%     \begin{tabular}{c|c|c|c}
				%     \hline
				%     \hline
				%     DNNs & Layers &  $J^{(\ell)}$ in hidden layers & \makecell{Learning\\ rate} \\
				%     \hline
				%     Edge-GCN & 4     & [128,128,128,128] & 0.002\\
				%     \hline
				%     Model-GNN & 3     & [32,32,32] & 0.002  \\
				%     \hline
				%     RGNN  & 3     & [32,32,32] & 0.002  \\
				%     \hline
				%     Indirect-GCN & 3 & [32,32,32] & 0.002 \\
				%     \hline
				
				%     2D-Gformer & 3     & [32,32,32] & 0.002 \\
				%     \hline
				%     3D-Gformer &4 &[128,128,128,128] & 0.005\\
				%     \hline

				%     DNNs & Layers &  \makecell{Dimension of $\mathbf{d}_{k}^{(\ell)}$ \\ in hidden layers} & \makecell{Learning\\ rate} \\
				%     \hline
				
				%     Vanilla Transformer &6 & \makecell{Encoder: [32$N$,32$N$,32$N$] \\ Decoder: [32$N$,32$N$,32$N$]}& 0.002\\
				%     \hline
				%     1D-Transformer & 3     & [32$N$,32$N$,32$N$] & 0.0005  \\
				%     \hline
				%     1D-Transformer-AN & 3     & [32$K$,32$K$,32$K$] & 0.002  \\
				%     \hline
				%     \hline
				%     \end{tabular}%
			
			% \end{threeparttable}
		%   %
		% \end{table}\vspace{-0.01mm}

	\vspace{-2mm}\subsection{Impact of the Simplification in the 2D-Gformer}\label{sec: ablation study}
	We validate by an ablation study that: (1) removing $\mathbf{U}^{\mathsf{Q}}$ and $\xi(\cdot)$ in \eqref{eq:2d-trans} can reduce the inference time without degrading the performance, and (2) removing $\mathbf{U}^{\mathsf{K}}$ or $\mathbf{U}^{\mathsf{V}}$ in \eqref{eq:upd-2d-trans-simp} entails performance degradation.
	
	Specifically, we evaluate the performance of the F-2D-Gformer whose attention sub-layer includes $\mathbf{U}^{\mathsf{Q}}$ and $\xi(\cdot)$. We also evaluate another two Graph Transformers, which are denoted as the ``2D-Gformer w/o $\mathbf{U}^{\mathsf{K}}$'' and ``2D-Gformer w/o $\mathbf{U}^{\mathsf{V}}$'', where $\mathbf{U}^{\mathsf{K}}$ and $\mathbf{U}^{\mathsf{V}}$ are respectively removed from the 2D-Gformer. All these DNNs are trained with 5,000 training samples generated in the scenarios with $N=16$, $K=8$, SNR = $\{5,10,20\}$ dB. The SE ratios and the inference time on CPU are listed in Table \ref{tab:ablation}.
	
	\begin{table}[!ht]
		\centering \vspace{-0.2mm}
		\caption{SE ratios and inference time of 2D-Gformer and its variants}\vspace{-2mm}
		\begin{tabular}{c|c|c|c|c}
			\hline\hline
			DNN & SNR=5dB & SNR=10dB & SNR=20dB & \makecell[c]{Time \\(ms)} \\
			\hline
			\textbf{2D-Gformer} & 99.46\% & 99.42\% & 99.12\% & 3.73  \\
			\hline
			\makecell[c]{F-2D-Gformer \\  }& 98.99\% & 98.54\% & 94.59\% & 4.27  \\
			\hline
			\makecell[c]{2D-Gformer \\ w/o $\mathbf{U}^{\mathsf{K}}$ }& 78.66\% & 69.45\% & 69.93\% & 3.54 \\
			\hline
			\makecell[c]{2D-Gformer\\ w/o $\mathbf{U}^{\mathsf{V}}$} & 96.72\% & 95.76\% & 97.42\% & 3.52 \\
			\hline\hline
		\end{tabular}%
		\label{tab:ablation}%
	\end{table} \vspace{-0.02mm}
	The results show that the proposed 2D-Gformer achieves the same or even higher SE ratio than the F-2D-Gformer especially for high SNRs with shorter inference time.
	This implies that removing $\xi(\cdot)$ enables more accurate modeling of the MUI.
	Besides, removing $\mathbf{U}^{\mathsf{K}}$ or $\mathbf{U}^{\mathsf{V}}$ leads to performance degradation in all scenarios, which validates that $\mathbf{U}^{\mathsf{K}}$ and $\mathbf{U}^{\mathsf{V}}$ should be preserved.
	
	\vspace{-2mm}\subsection{Learning Performance and Complexity}
\begin{figure}
		\centering
		\includegraphics[width=0.9\linewidth]{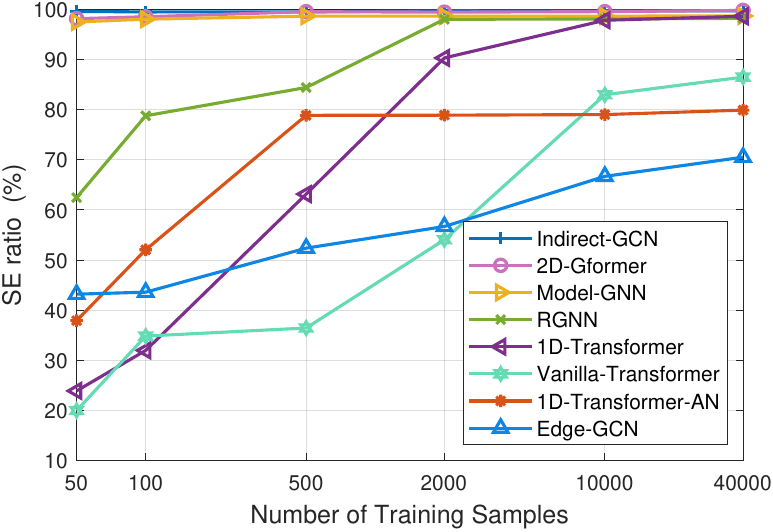}\vspace{-2mm}
		\caption{Learning performance of the baseband precoding. }
		\label{Sample Complexity}
	\end{figure}

	In Fig. \ref{Sample Complexity}, we compare the learning performance of the DNNs trained with different numbers of training samples for the baseband precoding.
	
	It can be seen that both the 2D-Gformer and  Model-GNN can achieve over 98\% SE ratio with only 50 training samples. The RGNN performs poorly with few samples because the parameterized function $q_{\mathsf{R}}(\cdot)$ needs to be learned with sufficient training samples. The 1D-Transformer only performs well with more than 10,000 samples, due to not exploiting the PE property of the precoding policy.  The Vanilla-Transformer, the 1D-Transformer-AN, and the Edge-GCN cannot achieve 90\% SE ratio even with 40,000 samples. This is because the Vanilla-Transformer
	does not satisfy any PE property, while the other two DNNs do not reflect the MUI in the processors.
	The Indirect-GCN performs the best performance, which however is only applicable to the baseband precoding in MISO systems.
	
	We have also evaluated the learning performance of the 2D-Gformer with a single head (i.e., $M_h=1$). The achieved SE ratio is almost the same as the 2D-Gformer with $M_h=32$. The results are not provided for a clear figure. This indicates that using multiple heads provides marginal performance gain for learning precoding.

	In Table \ref{Complex}, we provide the computational complexity in the inference phase in terms of running time and the number of floating-point operations (FLOPs), as well as the sample and space complexities in the training phase. The running time for inference is averaged over all the test samples. The FLOPs are measured by the THOP tools \cite{THOP}. The sample complexity and space complexity are respectively the number of training samples and the number of trainable parameters required by each DNN to achieve 98\% SE ratio. The results of the Edge-GCN and the 1D-Transformer-AN are not provided because they cannot achieve 98\% SE ratio even with 40,000 training samples.

	\begin{table}[htbp]
		\centering
		\begin{threeparttable}
			\caption{Inference and training complexity of each DNN}  \label{Complex}%
			\begin{tabular}{c|c|c|c|c|c}
				\hline
				\hline
				DNN  & \multicolumn{2}{c|}{ \makecell[c]{Inference time  \\ (CPU/GPU)(ms)}  } & FLOPs &   \makecell[c]{Training \\ samples  } &  \makecell[c]{Trainable \\ parameters  } \\
				\hline
				\textbf{2D-Gformer} & \quad 2.16 \quad & 2.18  &  0.086M   & 50 & 372 \\
				\hline
				Model-GNN & \quad 1.74 \quad & 1.69  &  0.033M   & 50 & 72 \\
				\hline
				Indirect-GCN & \quad 1.34 \quad & 2.08  &  0.01M\     & 1     & 12\\
				\hline
				RGNN  & \quad 5.82 \quad & 4.81 & 7M  & 2,000 & 0.028M \\
				\hline
				
				1D-Transformer &  \quad 4.01 \quad & 2.61 & 2M &10,000 & 0.323M\\
				
				\hline\hline
				\multicolumn{5}{l}{* ``M" means million for short.}\\
			\end{tabular}%
		\end{threeparttable}
	\end{table}
	
	It can be seen that the inference complexity and space complexity of the 2D-Gformer are higher than the two model-driven GNNs, but are lower than the RGNN and the 1D-Transformer.
	% the proposed 2D-Gformer has higher inference complexity and space complexity than the two model-driven GNNs but the complexities are much lower than the RGNN and 1D-Transformer. 
	The training complexity of the 1D-Transformer is the highest, owing to not satisfying the PE property of the antenna dimension. The Indirect-GCN has the lowest inference complexity and training complexity.
	
\begin{figure}
		\centering
		\includegraphics[width=0.9\linewidth]{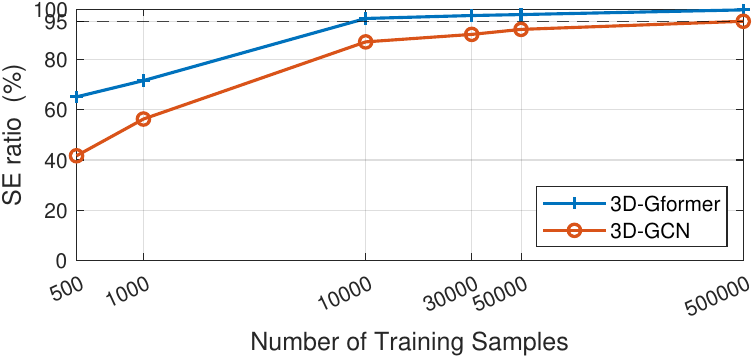}\vspace{-2mm}
		\caption{Learning performance of the hybrid precoding.}
		\label{fig:Performance_hybrid_precoding}
	\end{figure}

	In Fig. \ref{fig:Performance_hybrid_precoding}, we provide the learning performance of the 3D-Gformer for the hybrid precoding versus the number of training samples. We compare it with the 3D-GCN proposed in \cite{liu2023multidimensional} learning over the same heterogeneous graph as the 3D-Gformer, but do not compare with the two model-driven GNNs since they are not applicable. It can be seen that the 3D-Gformer requires much fewer training samples than the 3D-GCN to achieve the same performance (say 10,000/500,000 = 2\% to achieve the 95\% SE ratio).

	\vspace{-2mm}
	
	\vspace{-1mm}\subsection{Size-generalizability}\label{sec: sim-generalizability} \vspace{-1mm}
	To evaluate the generalizability to the numbers of users and antennas, each DNN for the baseband precoding is tested with the samples generated in scenarios with different values of $K$ and $N$ from the training samples without re-training.
	
\begin{figure}
		\centering
		\subfigure[Generalizability to the number of users, $N=16$.]{\includegraphics[width=0.9\linewidth]{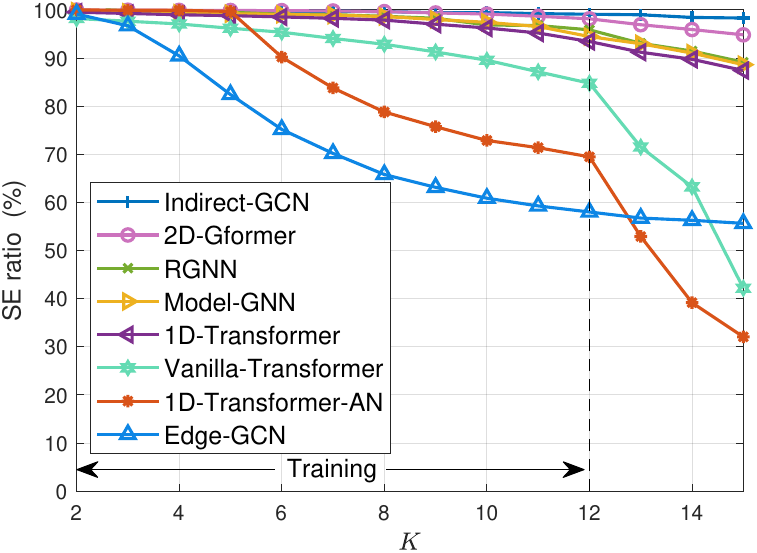}%
			\label{user Gen}}
		
		\subfigure[Generalizability to the number of antennas, $K=8$.]{   \includegraphics[width=0.9\linewidth]{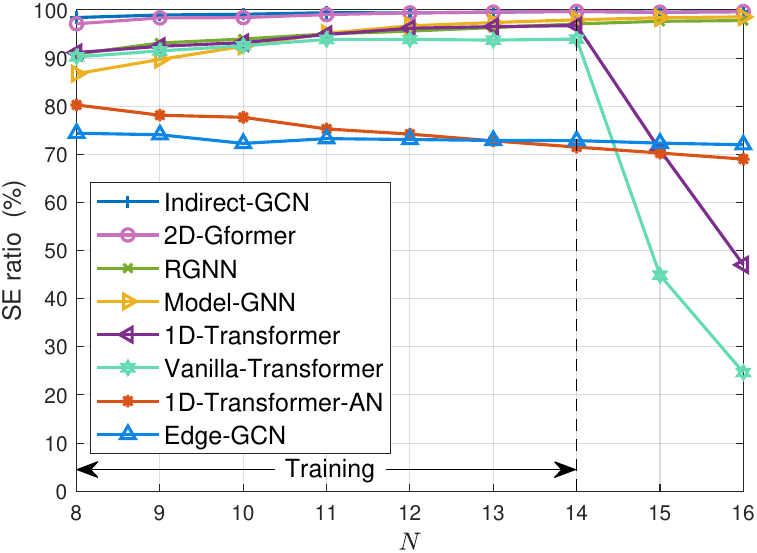}  \label{AN Gen}}
		\vspace{-3mm}
		\caption{Size-generalizability of DNNs for the baseband precoding.}
	\end{figure}

	In Fig. \ref{user Gen}, we evaluate the generalizability to the number of users. The training samples are generated in the scenarios with $N=16$ and different values of $K$, where $K$ is an integer generated from an exponential distribution with a mean value of 4 and truncated at 12. The test samples are generated in the scenarios with $N=16$ and $K$ randomly selected from a uniform distribution $\mathbb{U}(2,15)$. This setup ensures that the test set includes the samples with problem sizes not ``seen'' during training, while most training samples are generated at small problem sizes to reduce training complexity. Since the sizes of the weight matrices in the 1D-Transformer-AN depend on $K$,
	we set its input size as the maximum of the possible values of $K$ (i.e., 15) and pad the inputs with zeros when the number of users in a sample is less than 15.
	
	We can see that the SE ratios achieved by the 2D-Gformer, RGNN, Model-GNN, and 1D-Transformer decrease slower with $K$, indicating that they can be well-generalized to the number of users. However, the SE ratios achieved by the Vanilla-Transformer and 1D-Transformer-AN reduce rapidly when $K>12$, i.e., they are not generalizable to $K$. This is because they are not equivariant to the permutation of users. While both are without attention mechanism, the performance of the Edge-GCN gradually degrades with $K$ no matter if the number of users during testing has been ``seen'' in the training samples (i.e., $K \leq 12$) or not (i.e., $K > 12$), but the Indirect-GCN exhibits good size-generalizability to users, because no MUI needs to be avoided by the power allocation.
	
	\begin{figure}[htpb]
		\centering
		\subfigure[Generalizability to the number of users, $N=16$.]{\includegraphics[width=0.9\linewidth]{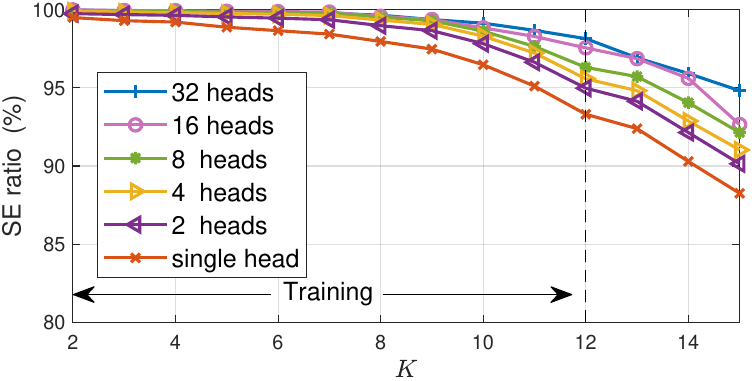}%
			\label{fig_first_case}}
		
		\subfigure[Generalizability to the number of antennas, $K=8$.]{\includegraphics[width=0.9\linewidth]{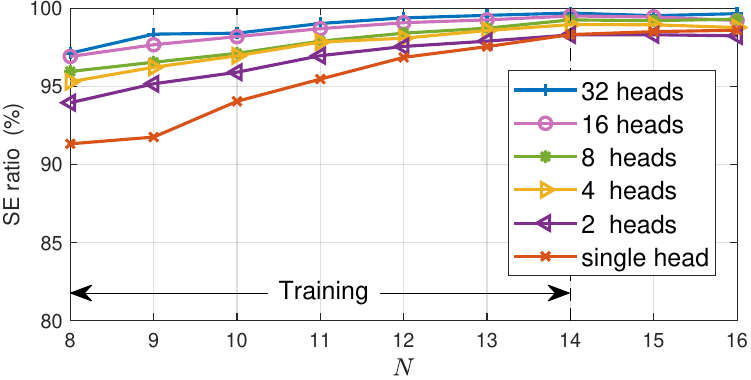}%
			\label{fig_second_case}} \vspace{-2mm}
		\caption{Size-generalizability of the 2D-Gformer with different $M_h$.}\label{fig:size-gen-diff-heads}
	\end{figure}
	
	In Fig. \ref{AN Gen}, we evaluate the generalizability to the number of antennas. The training samples are generated in the scenarios with $K=8$ and different values of $N$, where $N$ is an integer generated from an exponential distribution with a mean value of 10 and is truncated at 14. The test samples are generated in the scenarios with $K=8$ and $N$ randomly selected from $\mathbb{U}(8,16)$.
	Again, since the sizes of the weight matrices in the Vanilla-Transformer and 1D-Transformer depend on $N$, we set the input sizes of the two DNNs as the maximum of the possible values of $N$ (i.e., 16) and pad the inputs with zeros when the number of antennas in a sample is less than 16.
	
	It is shown that the 2D-Gformer, RGNN, Model-GNN, and Indirect-GCN can be well-generalized to the number of antennas. Both the 1D-Transformer-AN and Edge-GCN are generalizable to $N$, but do not perform well. The Vanilla-Transformer and 1D-Transformer cannot be generalized to $N$, since their performance degrades rapidly when $N>14$. This is because they do not satisfy the PE property of antennas.
	
	In Fig. \ref{fig:size-gen-diff-heads}, we show the impact of the number of heads on the size-generalizability of the 2D-Gformer, where several 2D-Gformers with different values of $M_h$ are trained and tested.
	It is shown that using more heads can improve the size-generalizability. However, when we further increase $M_h$ beyond 32, the size-generalizability improves little.
	
\begin{figure}[htpb]
		\centering
		\subfigure[Generalizability to the number of users, $N=32,N_\mathsf{RF}=8$.]{\includegraphics[width=0.9\linewidth]{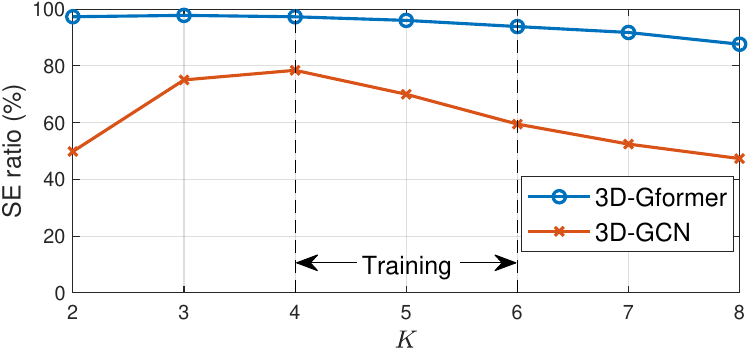}\label{fig:genK_hybird}}
		
		\subfigure[Generalizability to the number of antennas, $K=4,N_\mathsf{RF}=8$.]{\includegraphics[width=0.9\linewidth]{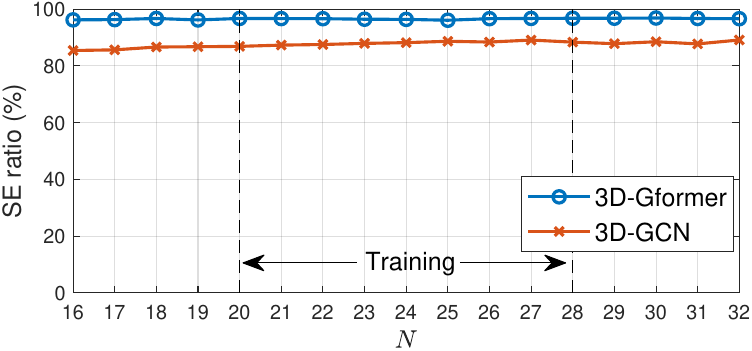}\label{fig:genN_hybird}}

		\subfigure[Generalizability to the number of RF chains, $N=32,K=4$.]{\includegraphics[width=0.9\linewidth]{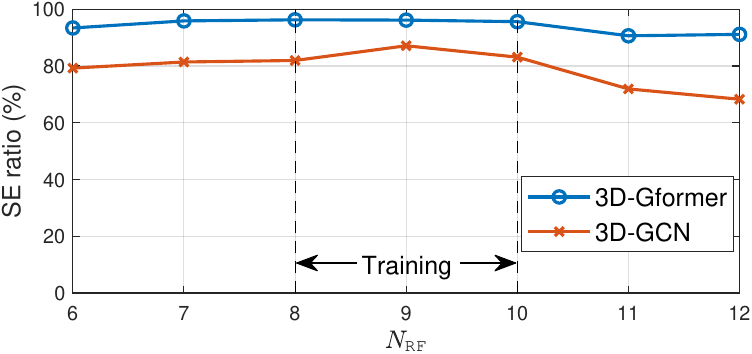}\label{fig:genNs_hybird}}
		\vspace{-2mm}
		\caption{Size-generalizability of DNNs for the hybrid precoding.}
		\label{fig:gen_Hybrid}
	\end{figure}

	In Fig. \ref{fig:gen_Hybrid}, we evaluate the size-generalizability of the 3D-Gformer and the 3D-GCN for learning the hybrid precoding policy. In Fig. \ref{fig:genK_hybird}, we provide the SE ratios under different values of $K$. The training samples are generated in the scenario with $N=32, N_\mathsf{RF}=8$ and $K$  randomly selected from $\mathbb{U}(4,6)$, and the test samples are generated in the scenario with $N=32, N_{\mathsf{RF}}=8$ and $K$ randomly selected from $\mathbb{U}(2,8)$.  In Fig. \ref{fig:genN_hybird}, we show the SE ratios under different values of $N$. The training samples are generated in the scenario with $K=4, N_\mathsf{RF}=8$ and $N$  randomly selected from $\mathbb{U}(20,28)$, and the test samples are generated in the scenario with $K=4, N_\mathsf{RF}=8$ and $N$ randomly selected from $\mathbb{U}(16,32)$. In Fig. \ref{fig:genNs_hybird}, we show the SE ratios under different values of $N_\mathsf{RF}$. The training samples are generated in the scenario with $N=32, K=4$ and $N_{\mathsf{RF}}$ randomly selected from $\mathbb{U}(8,10)$, and the test samples are generated in the scenario with $N=32, K=4$ and $N_{\mathsf{RF}}$ randomly selected  from $\mathbb{U}(6,12)$.
	
	It is shown that the 3D-Gformer can be well-generalized to the numbers of users, antennas, and RF chains, but the 3D-GCN cannot be generalized to the number of users. These results validate that the MUI should also be reflected by the processor for the generalizability to the number of users when learning the hybrid precoding policy.
	
	\section{Conclusions}\label{sec:conclusion}
	% In this paper,
	% we established the relation between the Transformer and GNNs for learning MU-MISO precoding.
	% We first identified the essential modules in the Transformer for learning the precoding policy.
	% By comparing the update equations of the Transformer and the GNNs, we found that the Transformer does not fully exploit the PE property of the precoding policy. Then, we proposed a 2D-Gformer for learning over a heterogeneous graph that can fully exploit the property. The update equation of the 2D-Gformer is simplified by comparing it with the update equations of GNNs, which helps reduce computational complexity.
	% %the Transformer, the Model-GNN, and RGNN are similar in both architecture and form. Understanding this relation can help us understand the Transformer for learning precoding and guide us to design a Transformer that learns precoding efficiently. Specifically, motivated by GNNs, we proposed a 2D-Gformer that satisfies the 2D-PE property for learning precoding.
	% Simulation results show that the 2D-Gformer can achieve or exceed the performance of model-based methods with few training samples. Additionally, the 2D-Gformer exhibits good size-generalizability to different problem sizes.
	
	In this paper, we answered the following two questions: is the Transformer the best choice for learning precoding policies and how should Transformers be
	designed to be size-generalizable? We found that the encoder-only Transformer
	without positional encoding is generalizable to the number of users for precoding by defining each user as a token, because the attention mechanism can reflect the MUI. However, the Transformer is less efficient than Graph Transformers since it cannot exploit the permutation property of precoding policies. To provide guidelines for designing Graph Transformers with the same PE properties to precoding policies, we established the relations between Transformers and the GNNs learning over heterogeneous graphs. We then proposed a 2D-Gformer for learning baseband precoding, and a 3D-Gformer for learning hybrid precoding, which are with matched permutation properties to the precoding policies. Simulation results showed that the 2D- and 3D-Gformers outperform the Transformers, and can be well size-generalized to different dimensions.

	%\appendices
	\begin{appendices} \numberwithin{equation}{section}
		\renewcommand{\thesectiondis}[2]{\Alph{section}}
		
		\section{Proof of Proposition \ref{prop:transformer-pe}}\label{proof:1D-PE}
		Denote $\mathbf{D}^{(\ell)} \triangleq [\mathbf{d}_1^{(\ell)},\cdots,\mathbf{d}_K^{(\ell)}]$. Then, from \eqref{eq:upd-transformer} all the token representations in the $\ell$-th layer can be updated as,
		\begin{equation}\label{eq:matrix_form upd transformer}
			\begin{aligned}
				&\mathbf{D}^{(\ell)}
				=\sigma\Bigg(\mathbf{W}^{\mathsf{F}} \mathbf{D}^{(\ell-1)}\\ &\quad\quad+\mathbf{W}^{\mathsf{F}}\mathbf{W}^{\mathsf{V}} \mathbf{D}^{(\ell-1)} \xi
				\Big(\big(\mathbf{W}^{\mathsf{K}} \mathbf{D}^{(\ell-1)}\big)^{\mathsf{T}} \big(\mathbf{W}^{\mathsf{Q}} \mathbf{D}^{(\ell-1)}\big)\Big) \Bigg)\\
				&\quad\quad\triangleq G(\mathbf{D}^{(\ell-1)}).
			\end{aligned}
		\end{equation}
		
		It has been proved in \cite{mehrabian2024joint,Transformer_PE} that $G(\cdot)$ satisfies the 1D-PE property in \eqref{1d-pe-UE}.
	Next, we prove that  $G(\cdot)$ does not satisfy the 2D-PE property in \eqref{2d-pe}.  Specifically, when $\mathbf{D}^{(\ell-1)}$ are permuted as  $\mathbf{\Pi}_{\mathsf{AN}}\mathbf{D}^{(\ell-1)}$, from \eqref{eq:matrix_form upd transformer} we have,
	\begin{equation}\label{eq:matrix_form upd transformer Pi_AN}
		\begin{aligned}
			&G(\mathbf{\Pi}_{\mathsf{AN}}\mathbf{D}^{(\ell-1)}) =\sigma\Bigg(\mathbf{W}^{\mathsf{F}} \mathbf{\Pi}_{\mathsf{AN}}\mathbf{D}^{(\ell-1)}+\mathbf{W}^{\mathsf{F}}\mathbf{W}^{\mathsf{V}}\cdot\\& \mathbf{\Pi}_{\mathsf{AN}}\mathbf{D}^{(\ell-1)} \xi \Big(\big(\mathbf{W}^{\mathsf{K}} \mathbf{\Pi}_{\mathsf{AN}}\mathbf{D}^{(\ell-1)}\big)^{\mathsf{T}} \big(\mathbf{W}^{\mathsf{Q}} \mathbf{\Pi}_{\mathsf{AN}}\mathbf{D}^{(\ell-1)}\big)\Big) \Bigg).
		\end{aligned}
	\end{equation}
	By comparing \eqref{eq:matrix_form upd transformer} and \eqref{eq:matrix_form upd transformer Pi_AN}, we can see that $G(\mathbf{\Pi}_{\mathsf{AN}}\mathbf{D}^{(\ell-1)}) \neq \mathbf{\Pi}_{\mathsf{AN}}G(\mathbf{D}^{(\ell-1)})$. Hence, the input-output relation of the update equation in \eqref{eq:matrix_form upd transformer} does not satisfy the 2D-PE property.
	
	By stacking $L$ layers with the input-output relation in each layer satisfying the 1D-PE property, it is not hard to prove that the input-output relation of the tailored Transformer satisfies the 1D-PE property in \eqref{1d-pe-UE} but does not satisfy the 2D-PE property.

	\section{Proof of Proposition \ref{lemma:1-revised}}\label{proof:lemma-revised}
	Since the element-wise activation function $\sigma(\cdot)$ does not affect permutation property, from  \eqref{eq:2d-trans} we can derive that
	\begin{eqnarray}\label{eq:left-revised}
		\mathbf{\Pi}_{\mathsf{AN}}\mathbf{d}_{\mathsf{E},k}^{(\ell)} \!\!&\!\!{=}\!\!&\!\!\sigma\Bigg(\mathbf{\Pi}_{\mathsf{AN}}\mathbf{U}^{\mathsf{F}} \Big(\mathbf{d}_{\mathsf{E},k}^{(\ell-1)}+\sum_{i=1}^{K}  \xi \Big(\big(\mathbf{U}^{\mathsf{Q}} \mathbf{d}_{\mathsf{E},k}^{(\ell-1)}\big)^{\mathsf{T}}\cdot \notag\\
		&& \big(\mathbf{U}^{\mathsf{K}} \mathbf{d}_{\mathsf{E},i}^{(\ell-1)}\big)\Big)\big(\mathbf{U}^{\mathsf{V}} \mathbf{d}_{\mathsf{E},i}^{(\ell-1)} \big)\Big)\Bigg) \notag\\
		\!\!&\!\!\overset{(a)}{=}\!\!&\!\!\sigma\Bigg(\mathbf{\Pi}_{\mathsf{AN}}\mathbf{U}^{\mathsf{F}} \Big(\mathbf{d}_{\mathsf{E},k}^{(\ell-1)}+\sum_{i=1}^{K}  \xi \Big(\big(\mathbf{\Pi}_{\mathsf{AN}}\mathbf{U}^{\mathsf{Q}} \mathbf{d}_{\mathsf{E},k}^{(\ell-1)}\big)^{\mathsf{T}}\cdot\notag\\
		&& \big(\mathbf{\Pi}_{\mathsf{AN}}\mathbf{U}^{\mathsf{K}} \mathbf{d}_{\mathsf{E},i}^{(\ell-1)}\big)\Big)\big(\mathbf{U}^{\mathsf{V}} \mathbf{d}_{\mathsf{E},i}^{(\ell-1)} \big)\Big)\Bigg),
	\end{eqnarray}
	where $(a)$ holds because $\mathbf{a}^{\mathsf{T}}\mathbf{b}=\mathbf{a}^{\sf T}\mathbf{\Pi}_{\sf AN}^{\sf T}\mathbf{\Pi}_{\mathsf{AN}}\mathbf{b}= (\mathbf{\Pi}_{\mathsf{AN}}\mathbf{a})^{\mathsf{T}}(\mathbf{\Pi}_{\mathsf{AN}}\mathbf{b})$ and $\mathbf{\Pi}_{\sf AN}^{\sf T}\mathbf{\Pi}_{\mathsf{AN}}=\mathbf{I}$.
	
	When $\mathbf{D}^{(\ell-1)}$ are permuted as  $\mathbf{\Pi}_{\mathsf{AN}}\mathbf{D}^{(\ell-1)}$, the right-hand side of \eqref{eq:pi-upd-2D-Gformer-revised} can be obtained from \eqref{eq:2d-trans} as,
	\begin{eqnarray}\label{eq:right-revised}
		&&F(\mathbf{\Pi}_{\mathsf{AN}}\mathbf{d}_{\mathsf{E},1}^{(\ell-1)},\cdots,\mathbf{\Pi}_{\mathsf{AN}}\mathbf{d}_{\mathsf{E},K}^{(\ell-1)})\notag\\ \!\!&\!\!=\!\!&\!\!\sigma\Bigg(\mathbf{U}^{\mathsf{F}} \Big(\mathbf{\Pi}_{\mathsf{AN}}\mathbf{d}_{\mathsf{E},k}^{(\ell-1)}+\sum_{i=1}^{K}  \xi \Big(\big(\mathbf{U}^{\mathsf{Q}} \mathbf{\Pi}_{\mathsf{AN}}\mathbf{d}_{\mathsf{E},k}^{(\ell-1)}\big)^{\mathsf{T}} \cdot\notag\\
		&&\big(\mathbf{U}^{\mathsf{K}} \mathbf{\Pi}_{\mathsf{AN}}\mathbf{d}_{\mathsf{E},i}^{(\ell-1)}\big)\Big)\big(\mathbf{U}^{\mathsf{V}} \mathbf{\Pi}_{\mathsf{AN}}\mathbf{d}_{\mathsf{E},i}^{(\ell-1)} \big)\Big)\Bigg) \notag\\
		\!\!&\!\!\overset{(a)}{=}\!\!&\!\!\sigma\Bigg(\mathbf{U}^{\mathsf{F}}\mathbf{\Pi}_{\mathsf{AN}} \Big(\mathbf{d}_{\mathsf{E},k}^{(\ell-1)}+\sum_{i=1}^{K}  \xi \Big(\big(\mathbf{U}^{\mathsf{Q}} \mathbf{\Pi}_{\mathsf{AN}}\mathbf{d}_{\mathsf{E},k}^{(\ell-1)}\big)^{\mathsf{T}} \cdot\notag\\
		&&\big(\mathbf{U}^{\mathsf{K}} \mathbf{\Pi}_{\mathsf{AN}}\mathbf{d}_{\mathsf{E},i}^{(\ell-1)}\big)\Big)\big(\mathbf{\Pi}_{\mathsf{AN}}^{\mathsf{T}}\mathbf{U}^{\mathsf{V}} \mathbf{\Pi}_{\mathsf{AN}}\mathbf{d}_{\mathsf{E},i}^{(\ell-1)} \big)\Big)\Bigg),
	\end{eqnarray}
	where $(a)$ holds due to the fact that $\mathbf{\Pi}_{\mathsf{AN}}\mathbf{\Pi}_{\mathsf{AN}}^{\mathsf{T}}=\mathbf{I}$.
	
	We can see that if $\mathbf{\Pi}_{\mathsf{AN}}\mathbf{U}^{\mathsf{F}}=\mathbf{U}^{\mathsf{F}}\mathbf{\Pi}_{\mathsf{AN}}$, $ \mathbf{\Pi}_{\mathsf{AN}}\mathbf{U}^{\mathsf{Q}}=\mathbf{U}^{\mathsf{Q}}\mathbf{\Pi}_{\mathsf{AN}}$, $\mathbf{\Pi}_{\mathsf{AN}}\mathbf{U}^{\mathsf{K}}=\mathbf{U}^{\mathsf{K}}\mathbf{\Pi}_{\mathsf{AN}}$, and $\mathbf{U}^{\mathsf{V}}=\mathbf{\Pi}_{\mathsf{AN}}^{\mathsf{T}}\mathbf{U}^{\mathsf{V}}\mathbf{\Pi}_{\mathsf{AN}}$, then \eqref{eq:left-revised} equals \eqref{eq:right-revised}.
	Since $\mathbf{\Pi}_{\mathsf{AN}}\mathbf{\Pi}^{\mathsf{T}}_{\mathsf{AN}}=\mathbf{I}$, the fourth equality can be re-expressed as $\mathbf{\Pi}_{\mathsf{AN}}\mathbf{U}^{\mathsf{V}}=\mathbf{U}^{\mathsf{V}}\mathbf{\Pi}_{\mathsf{AN}}$ by left-multiplying $\mathbf{\Pi}_{\mathsf{AN}}$ on both sides.
	
\end{appendices}

\ifCLASSOPTIONcaptionsoff
\newpage
\fi

% \end{thebibliography}
\bibliographystyle{IEEEtran}
\bibliography{bibtex/bib/IEEEabrv,bibtex/bib/IEEEexample}

\end{document}